\pdfoutput=1
\documentclass[11pt]{article}

\usepackage{jheppub}
\usepackage{amsmath}
\usepackage{verbatim}
\usepackage{amssymb}
\usepackage{inputenc}
\usepackage{textcomp}
\usepackage{appendix}
\usepackage{mathtools}
\usepackage{multirow}
\usepackage{mathrsfs}
\usepackage{stmaryrd}
\usepackage{graphicx}
\usepackage{subcaption}
\usepackage{float}
\usepackage{blindtext}
\usepackage{tcolorbox}
\usepackage{empheq}
\usepackage[active]{srcltx}
\usepackage{tikz}
\usepackage{lmodern}
\usepackage{braket}
\usetikzlibrary{decorations.pathmorphing}
\tikzset{snake it/.style={decorate, decoration=snake}}
\usepackage{fancybox}
\newcommand{\be}[1]{\begin{equation}\label{#1} }
\newcommand{\ee}{\end{equation}}
\newcommand{\bea}[1]{\begin{eqnarray}\label{#1} }
\newcommand{\eea}{\end{eqnarray}}
\newcommand{\bes}{\begin{subequations}}
\newcommand{\ees}{\end{subequations}}

\newcommand{\mthO}{\mathcal{O}}
\newcommand{\mthA}{\mathcal{A}}
\usepackage{pifont}

\title{\bf Operator growth and Krylov Complexity in Bose-Hubbard Model}
\author[a]{Arpan Bhattacharyya,}
\author[b,c]{Debodirna Ghosh,}
\author[a,d]{Poulami Nandi}
\affiliation[a]{\it Indian Institute of Technology, Gandhinagar, Gujarat-382355, India}
\affiliation[b]{\it Institute of Mathematical Sciences, IV Cross Road, C.I.T. Campus, Taramani, Chennai 600113, India}
\affiliation[c]{\it Homi Bhabha National Institute, Training School Complex, Anushakti Nagar, Mumbai 400094, India}
\affiliation[d]{\it David Rittenhouse Laboratory, University of Pennsylvania, 209 S.33rd Street, Philadelphia, PA
19104, USA}
\emailAdd{abhattacharyya@iitgn.ac.in}
\emailAdd{debodirna@imsc.res.in}
\emailAdd{pnandi@sas.upenn.edu}
\abstract{We study Krylov complexity of a one-dimensional Bosonic system, the celebrated Bose-Hubbard Model. The Bose-Hubbard Hamiltonian consists of interacting bosons on a lattice, describing ultra-cold atoms. Apart from showing superfluid-Mott insulator phase transition, the model also exhibits both chaotic and integrable (mixed) dynamics depending on the value of the interaction parameter. We focus on the three-site Bose Hubbard Model (with different particle numbers), which is known to be highly mixed. We use the Lanczos algorithm to find the Lanczos coefficients and the  Krylov basis. The orthonormal Krylov basis captures the operator growth for a system with a given Hamiltonian. However, the Lanczos algorithm needs to be modified for our case due to the instabilities instilled by the piling up of computational errors. Next, we compute the Krylov complexity and its early and late-time behaviour. Our results capture the chaotic and integrable nature of the system.  Our paper takes the first step to use the Lanczos algorithm non-perturbatively for a discrete quartic bosonic  Hamiltonian without depending on the auto-correlation method.}
\makeatletter
\begin{document}
\maketitle
\flushbottom



\section{Introduction}
 Quantum mechanics is a staple ingredient in the playground of every theoretical physicist and is also responsible for most modern developments in natural sciences. A basic idea during our introductory lessons in quantum mechanics is the evolution of a simple operator $\mathcal{O}$ with time $t$ for a system governed by a  Hamiltonian $H$. In the Heisenberg picture, the simple operator $\mathcal{O}$ at $t=0$ grows in the Hilbert Space into a complex operator $\mathcal{O}(t)$ through $e^{iHt}\mathcal{O}e^{iHt}$. This growth quantifies the spread of local information supported by a single or small number of sites to the entire system. The nature of the operator growth depends on the system's Hamiltonian. An operator, for example, will grow into an integrable and non-integrable system in distinct ways. Using the  Baker-Campbell-Hausdorff formula, the operator growth is expressed in terms of complicated nested commutators $[H,[H,\cdots[H,\mathcal{O}]]]$. We soon realize that the evaluation becomes rapidly difficult to perform with each level of the nested commutator. Nevertheless, this simple notion of operator growth plays a massive role in diverse areas of physics. The `dissipative' behaviour of the operator growth is recently found to be connected to the thermalization of a given system \cite{vonKeyserlingk:2017dyr,Nahum:2017yvy,Khemani:2017nda,Rakovszky:2017qit,Gopalakrishnan:2018wyg}. 

 \medskip
 \noindent
Another common way of describing the operator growth in the space-time picture is by the out-of-time-ordered (OTOC) commutator \cite{1969JETP...28.1200L,Xu:2018dfp,Nahum:2017yvy,vonKeyserlingk:2017dyr,Gu:2021xaj}. The classical chaos makes the OTOC grow exponentially, also known as the Lyapunov exponent. The interesting conjecture is that the Lyapunov exponent is bounded from above \cite{Maldacena:2015waa} for a few strongly coupled large-N models (the Sachdev-Ye-Kitaev (SYK) model, for example, having a dual in the gravity side) \cite{Maldacena:2016hyu,Kitaev15,PhysRevLett.70.3339}. However, the bound on the Lyapunov exponent is not seen universally, especially for generic, non-integrable Hamiltonians outside of the semi-classical, large-N limit \cite{Khemani:2017nda,PhysRevE.89.012923,Xu:2018xfz,Xu:2018dfp}. 

   \smallskip
  \noindent
 A general theory of operator growth under generic, non-integrable Hamiltonian dynamics is recently given by Parker et al \cite{Parker:2018yvk}. The authors argued that, 
 with each increasing level of the nested commutators in the operator growth, the growing operators increase exponentially in time. The exponential size of the computation enables the system to be essentially treated as a thermodynamic entity, thus, emerging a statistical picture. This gave rise to the idea of governing the operator growth with some universal properties, even for non-integrable systems in arbitrary dimensions.
The hypothesis for universal operator growth in \cite{Parker:2018yvk}  is based on the notion of a recursive technique, the Lanczos algorithm. Starting with a test operator $|\mathcal{O}(t=0))$, the Lanczos algorithm iteratively generates a set of orthonormal basis operators. This basis of orthonormal operators is called the Krylov basis $|\mathcal{O}_{n})$, which can have fewer elements than the dimension of the Hilbert space, depending on the dynamics and the choice of the initial operator. Along with the basis operators $|\mathcal{O}_{n})$ (where $n=0,1,\cdots,$ denotes the dimension of Krylov space), a set of coefficients $\{b_{n}\}$ is generated, known as Lanczos coefficients. The time evolved operator $|\mathcal{O}(t))$ is expanded in terms of the Krylov basis $|\mathcal{O}_{n})$ and  time-dependent coefficients $\phi_n(t)$, which obeys  discrete Schr\"{o}dinger equations \cite{Parker:2018yvk}.  The operator dynamics is encoded in these time-dependent coefficients, which can also be thought of as wavefunctions distributed over the Krylov basis. The operator growth can be understood as the delocalization of an initial operator which we choose to be  $|\mathcal{O}_{0})$ over the Kyrlov basis  with time, and the hopping amplitudes 
 are given by the Lanczos coefficients.
 
\smallskip
 \noindent
 The concept of various notions of complexity follows naturally from the above observation. For example, Parker et.al \cite{Parker:2018yvk} have introduced the notion of Krylov complexity, ({\color{black} or K-complexity) as a measure of quantum chaos in the thermodynamic limit, that depends on the local Hamiltonian, a specific choice of an inner product, and the choice of the reference operator. K-Complexity of a generic operator in a finite-dimensional system size exhibits a specific profile with time which was first shown in \cite{Barbon:2019wsy} and then in \cite{Rabinovici:2020ryf}. With an aim to find the chaotic nature of the system, we perform a numerical analysis to find the profile of K-Complexity at all time scales, including the scrambling phase and the late time phase, and compare our results to those of \cite{Barbon:2019wsy, Rabinovici:2020ryf}.}

\smallskip
 \noindent The Lanczos coefficients ${b_{n}}$, in a quantum chaotic system, are hypothesized to grow fast. The maximum possible growth varies linearly with $n$ \cite{Parker:2018yvk}, modified by a logarithmic correction in local $(1+1$)-dimensional lattice systems. This is in contrast, to the integrable models, {\color{black} \cite{Barbon:2019wsy} where the Lanczos coefficients grow as $b_{n}\sim \alpha n^{\delta}$, where $0< \delta < 1$. This behaviour of the Lanczos coefficients $b_{n}$, implies that the K-complexity grows exponentially in time in chaotic systems and powerlike in an integrable system.} In recent times this has motivated lots of studies. Interested readers are referred to some of these references, which are by no means exhaustive \cite{Barbon:2019wsy,Jian:2020qpp,DymarskyPRB2020,PhysRevLett.124.206803,Yates2020,Rabinovici:2020ryf,Rabinovici:2021qqt,Yates:2021lrt,Yates:2021asz,Dymarsky:2021bjq,Noh2021,Trigueros:2021rwj,https://doi.org/10.48550/arxiv.2207.13603,Fan_2022,Kar_2022,https://doi.org/10.48550/arxiv.2109.03824,PhysRevE.106.014152,https://doi.org/10.48550/arxiv.2204.02250,Bhattacharjee_2022,Bhattacharjee_2022a,Du:2022ocp, Banerjee:2022ime,
https://doi.org/10.48550/arxiv.2205.12815,https://doi.org/10.48550/arxiv.2207.05347,H_rnedal_2022,https://doi.org/10.48550/arxiv.2208.13362,Bhattacharjee:2022lzy,Rabinovici:2022beu,Alishahiha:2022anw, Avdoshkin:2022xuw,Kundu:2023hbk,Rabinovici:2023yex,Zhang:2023wtr,Nizami:2023dkf,Hashimoto:2023swv,Nandy:2023brt,Balasubramanian:2022tpr, Balasubramanian:2021mxo, Caputa2022PRB, Caputa:2022yju, https://doi.org/10.48550/arxiv.2208.10520,Erdmenger:2023shk,Bhattacharya:2023zqt,Chattopadhyay:2023fob,Pal:2023yik}.\par
Motivated by all these, we have initiated a study of operator growth for the Bose-Hubbard model in this paper. It has found important application in the field of ultracold atom physics \cite{jaksch1998cold}. This model exhibits Superfluidity-Mott insulator phase transition \cite{sachdev1999quantum}, which has been observed experimentally with ultracold atoms in optical lattice \cite{Sengupta:2005}. More importantly, this model provides us with a rich parameter space, tuning by which it can behave as chaotic or non-chaotic (integrable) systems \cite{PhysRevA.74.033601,Viscondi_2011,PhysRevA.101.053604}. Although for the system with fewer degrees of freedom, this kind of mixed behaviour can be expected \cite{FEINGOLD1983433,PhysRevA.30.509,PhysRevA.35.1464}, but for a generic quantum many-body system mixed systems are less common \cite{Nakerst:2022prc}. So this model  provides us with an ideal  playground to study operator growth as well as the associated Krylov complexity for both the chaotic and integrable regime and make a comparison.  In \cite{Nakerst:2022prc}, the authors have considered a classical limit by parametrizing the interaction by $\beta=U\,N$, where $U$ is the on-site interaction and the particle number is $N$, and then showed that in the limits $\beta\rightarrow 0$ and $\beta\rightarrow \infty$, the three-site Bose-Hubbard model possess integrable nature due to the vanishing of the only positive Lyapunov exponent $\lambda_{\text{max}}.$ Also, this model presents us with an excellent opportunity to study the operator growth and the Krylov complexity for a quantum-many body system consisting of bosonic degrees of freedom system containing quartic interaction terms using the Lanczos algorithm non-perturbatively. To the best of our knowledge, in the past, only limited studies of this kind have been made for non-quadratic systems \footnote{Non-Gaussian RMT  are studied in \cite{Balasubramanian:2022dnj,Erdmenger:2023shk}, but they are $0+1D$ systems. For SYK see \cite{Parker:2018yvk,Jian:2020qpp}. However, they are fermionic systems. For studies in JT gravity see gravity see \cite{Kar_2022} and for holographic CFT see \cite{Dymarsky:2021bjq}. Also, in \cite{Camargo:2022rnt} a study of Krylov complexity has been made for $\lambda \phi^4$ upto first order in $\lambda.$ Unlike our case, authors have used the auto-correlation method \cite{Parker:2018yvk,Dymarsky:2021bjq} to compute Lanczos coefficients and they work in the continuous limit. Also for a study of the Krylov complexity for Calabi-Yau quantum mechanics please refer to \cite{Du:2022ocp}. But this is a single particle quantum mechanical system.}.\par
The paper is organized as follows. In Sec.~\ref{BHmod}, we introduce the necessary details of the celebrated Bose-Hubbard model, used in the context of our current analysis. In Sec.~\ref{KrySp}, we dive into the details of the Krylov basis and spanning of the Krylov space, which is significantly different from the Hilbert space. Then, in Sec.~\ref{Numerical}, we revisit the Lanczos algorithm. Even after being a powerful technique, Lanczos method suffers terribly from instabilities and errors rounding up in the process. This lead to this Sec.~\ref{Numerical}, where we discuss the resolution to apply the modified version of the Lanczos method. Our results of the operator growth for the Bose-Hubbard model is presented in Sec.~\ref{Results} and that of the Krylov complexity is presented in Sec.~\ref{Results1}. Finally we end with a summary of our results and future directions in Sec.~\ref{discussion}. {\color{black}We also have incorporated Appendices~\eqref{A}, \eqref{A2} and \eqref{B} explaining more details of our numerical analysis.} 
\section{Bose-Hubbard model} \label{BHmod}
In this section, we start by summarising the necessary details of the one-dimensional Bose-Hubbard model. The Bose-Hubbard model consists of interacting bosons placed on a lattice. The model rose to prominence to describe the dynamics of ultracold bosonic atoms in an optical lattice, with the system parameters being completely controllable by laser light \cite{jaksch1998cold}.
We restrict the model to open-boundary chains of length $M$ with a vanishing chemical potential. The Hamiltonian is given by
\begin{eqnarray}\label{hamlt}
\label{BHHam}H = -J \sum_{<i,j>}^{M-1} (a_{i}^{\dag}a_{j}+ a_{j}^{\dag}a_{i}) + \frac{U}{2}\sum_{i=1}^{M} \hat{n}_{i}(\hat{n}_{i}-1)
\end{eqnarray}
where $<i,j> \equiv <j,i>$ denotes the summation over adjacent sites $(j=i\pm1)$, $a_{i}^{\dag}$ and $a_{i}$ are the bosonic creation and annihilation operators on the site $i$ (with the site $i$ running from 1 to the site number). Also, $\hat{n}_{i}=a_{i}^{\dag}a_{i}$ is the number operator that counts the number of particles on that site $i$. Here, $M$ denotes the site number, and $N$ is the particle number. In the first term, the summation runs up to $M-1$ when we consider the fixed boundary condition, such that particles are to be arranged in a linear chain with non-interacting walls at the end. Another possible boundary condition is to put the particles in a circle with the $M+1$-th site being identified as the first site. In this periodic boundary condition, the summation over the first term runs up to $M$. $J$ is the tunnelling coefficient which signifies the energy gained due to the particles hopping from one site to its neighbouring site. In this paper, we will only use open boundary condition. The first term proportional to $J$ is the kinetic term of the Hamiltonian. On the other hand, the parameter $U$ characterizes the two-particle on-site repulsive interaction strength.   A point of importance is that the parameters $J$ and $U$ can be adjusted by various means. The model with a finite value of chemical potential shows two different phases depending on the ratio of the two parameters $\Lambda \equiv \frac{U}{J}$ measured in terms of the energies. For $\Lambda << 1$, the phase is identified by weak coupling and strong hopping, giving rise to the super-fluid phase. Additionally, $\Lambda >>1$ with strong coupling and weak hopping characterizes the Mott-insulator phase \cite{Marvin}.

\smallskip
\noindent  
The Hamiltonian has a $U(1)$ symmetry associated with the conservation of a total number of particles. The symmetry group of the open boundary Hamiltonian can therefore be written as a $U(1)$. For closed boundary conditions, the symmetry group of the Hamiltonian becomes $U(1)\otimes D_{M}$, where the symmetry $D_{M}$ is the combination of translation and reflection symmetries. The dimension of the Hilbert space $\mathcal{H}$ with the total number of atom $N$ and number of sites $M$ is found to be
\begin{eqnarray}
D=\frac{(N+M-1)!}{N!(M-1)!}\,.
\end{eqnarray}
The dimension $D$ grows with the number of particles $N$ as $D\sim N^{M+1}$ for a fixed $M$. A classical limit exists in the Bose-Hubbard model when the particle number $N\to \infty$ with the site number $M$ is kept fixed. The classical limit is similar to the discreet non-linear Schrodinger equation. The interaction is usually parametrized in the classical limit by $\mathfrak{\beta}=UN$. The model in the $\beta=0$ and $\beta=\infty$ limits are integrable, thus, analytically solvable. For the site number $M=2$, the model is solvable for all values of $\beta$. Otherwise, for $M\geq 3$ and finite $\beta$, the Bose-Hubbard model is non-integrable \cite{Nakerst:2022prc} \footnote{To understand aspects of integrability and chaotic behaviours of the 1D Bose Hubbard model please also see the references within \cite{Nakerst:2022prc}.}. However, for $M=3$, the model shows neither strongly chaotic nor integrable behaviour. It is said to exhibit highly mixed behaviour. Our findings in Sec~\ref{Results}, also confirm the said behaviour. In the following sections, we analyze the Bose-Hubbard model for site number $M=3$ with different values of the particle number $N$ and $\Lambda \equiv \frac{U}{J}$. It is possible to generalize our results for higher particle and site numbers. However, the computations grow excessively expensive and are thus limited by the computational power. We will study the time evolution of an operator with a fixed Hamiltonian Eq.\eqref{hamlt} in Krylov space. We use the Lanczos algorithm to find the appropriate Krylov basis and the corresponding Lanczos coefficients. To the best of our understanding, this is the first time the Lanczos algorithm (appropriately modified, discussed in later sections) is being used for a bosonic lattice system.

\section{Revisiting the Krylov Space}
In this section, we review the details of the Krylov Basis and its relation to the Krylov space. The Krylov basis is an ordered, orthonormal set of base kets that contains information about the time evolution of an operator.
We begin by choosing an appropriate basis associated with the Hilbert space $\mathcal{H}$ with dimension $dim(\mathcal{H})=D$. The associated Hilbert space of linear operators $L(\mathcal{H})\equiv\hat{\mathcal{H}}$ that acts on $\mathcal{H}$, is of dimension $D^2$. Next, we choose a system of $N$ particles in $M$ sites governed by the Hamiltonian ($H$) defined in Eq.\eqref{hamlt} and an operator $\mathcal{O}$ ($H,\mathcal{O} \in \hat{\mathcal{H}} $) \cite{Rabinovici:2020ryf}. A natural choice of basis for  the Bose-Hubbard Hamiltonian is the base ket with the occupation numbers in each site, i.e.\ $\ket{{ n_{1},n_{2},\cdots,n_{M}}}$, which are eigenstates of number operators $\hat{n_{i}}$  
\begin{eqnarray}
\hat{n_{i}}\,\ket{{ n_{1},n_{2},\cdots ,n_{M}}} = n_{i}\,\ket{{ n_{1},n_{2},\cdots ,n_{M}}}.
\end{eqnarray}
Here, $n_{i}\geq 0$ and $\sum_{i=1}^{N}\,n_{i}=N$. One can construct all the basis vectors with appropriate $N$, and $M$ following the rules mentioned in \cite{Zhang_2010}.

\smallskip\noindent
 Next, we choose an initial operator $\mathcal{O}\in \hat{\mathcal{H}} $ and evolve it in time. The operator $\mathcal{O}$ gradually explores a subspace of the space of operators of the given system during its evolution. This subspace is called the Krylov space. The orthonormal Krylov basis $\mathcal{K}$ and the structure of the Krylov space can be analyzed using the powerful algorithm known as the Lanczos algorithm \cite{Parker:2018yvk}. The Lanczos algorithm also captures the rate at which a typical operator span the Krylov space. The Krylov basis $\mathcal{K}$  depends on the choice of the initial operator, and the dynamics of the system may not expand the entire Hilbert space $\mathcal{H}$. We use the smooth-ket notation $|\mathcal{O})$ to denote the element of the operator space corresponding to an operator $\mathcal{O}$  throughout the paper. In the following paragraph, we explain the notions of the Krylov Basis and Krylov space in the context of an operator growth $\mathcal{O}(t)$. \par

In the Heisenberg picture, the time evolution of $|\mathcal{O})$ is given by
\begin{eqnarray}\label{opexp}
\Big |\mathcal{O}(t)\Big)= e^{iHt}|\mathcal{O})e^{-iHt}= e^{i\mathcal{L}t}|\mathcal{O}).
\end{eqnarray}
Here, $\mathcal{L}$ is the Liouvillian operator defined by $\mathcal{L}|\mathcal{O})=\Big|[H,\mathcal{O}]\Big)$ with $H$ being the system Hamiltonian Eq.\eqref{hamlt}. Thus, using the Baker-Campbell-Hausdorff formula in Eq.\eqref{opexp},  the Krylov space $\mathcal{H}_{\mathcal{O}}$ associated with the operator $\mathcal{O}$,  can be defined as the linear span of all nested commutators of the Hamiltonian with the operator:
 \begin{eqnarray} \label{eq1}
 \mathcal{H}_{\mathcal{O}}= \text{span}\Big\{\mathcal{L}^{n}\,\mathcal{O}\Big\}^{+\infty}_{n=0}=\text{span}\{\mathcal{O},[H,\mathcal{O}],[H,[H,\mathcal{O}]],\cdots\}.
\end{eqnarray}
The dimension of the Krylov space $dim( \mathcal{H}_{\mathcal{O}})\equiv K$, is less than the dimension of the operator space $\hat{\mathcal{H}}$, $(dim(\hat{\mathcal{H}})=D^{2}$). For most systems, even though this set mentioned in Eq.(\ref{eq1}) has infinite elements, only a few of them are linearly independent. One can do a spectral decomposition \cite{Rabinovici:2020ryf} of $\mathcal{O}$ in the energy basis in order to obtain a precise result :
{\color{black}
\begin{eqnarray}\label{specdec}
|\mathcal{O})= \sum_{a,b}^{D} O_{ab}\ket{E_{a}}\bra{E_{b}}
\end{eqnarray}
where $\{E_{a},\ket{E_{a}}\}_{a=1}^{D}$ are the eigenvalues and eigenstates (respectively) of $H$. It can be shown following \cite{Rabinovici:2020ryf} that the elements of the sum in \eqref{specdec} are the eigenstates of the Liouvillian $\mathcal{L}$ :
\begin{eqnarray}
  \mathcal{L} |\omega_{ab})=  \omega_{ab}|\omega_{ab}),
\end{eqnarray}
with $|\omega_{ab})\equiv \ket{E_{a}}\bra{E_{b}}$, and ({\it phases}) $\omega_{ab}=E_{a}-E_{b}$. In general, the Krylov dimension $K$ is equal to the number of nonvanishing projections of $\mathcal{O}$ over the eigenspaces of the Liouvillian \cite{Rabinovici:2020ryf}. Therefore, the upper bound of the dimension $K$ is given by :
\begin{eqnarray}
\label{Kbasisbound}K\leq D^{2}-D+1.
\end{eqnarray}
This consists only of the unavoidable degeneracies, i.e,  the zero eigenvalue phase $\omega_{aa}=0$ and has the degeneracy of $D$. The upper bound $K= D^{2}-D+1$ is only saturated for cases with no degeneracies other than the diagonal null phases $\omega_{aa}$. The dimension of the Krylov space decreases with the increase of degeneracies in the Liouvillian spectrum. }

\smallskip
\noindent
Ideally, a typical {\color{black}Liouvillian operator in a chaotic system has no degeneracies other than the null phases} and is known to saturate \cite{Rabinovici:2020ryf} the upper bound of Eq.\eqref{Kbasisbound} while for integrable systems, the Krylov dimension $K$ is substantially smaller than the bound.

\section{Details of the Lanczos method}\label{KrySp}
After reviewing the necessary details of the Krylov space, we are now ready to construct it for our system. In this section, we review the Lanczos algorithm \cite{Lanczos:1950zz,vish123} to construct the Krylov Basis and, subsequently, the Krylov Space before discussing its shortcomings. \par
Given a certain inner product in the operator space, one can construct an orthonormal basis for Krylov space associated with an operator $\mathcal{O}$, with a fixed Hamiltonian $H$. This operation is done by the Lanczos algorithm, which is another version of the Gram-Schmidt orthogonalization procedure. In this paper, we make use of the infinite-temperature inner product, namely the Frobenius inner product \footnote{There are other finite-temperature choices of the inner product \cite{Parker:2018yvk,vish123,https://doi.org/10.48550/arxiv.2205.12815} which can be reduced to the Frobenius inner product at infinite temperature.} :
\begin{eqnarray}
\label{defnorm}
(\mathcal{O}_{1}|\mathcal{O}_{2}) =\frac{1}{D}\text{Tr}\Big[\mathcal{O}_{1}^{\dag}\mathcal{O}_{2}\Big],\,\,\,\,\, ||\mathcal{O}||^2=(\mathcal{O}|\mathcal{O})=\frac{1}{D}\text{Tr}\Big[\mathcal{O}^{\dag}\mathcal{O}\Big],
\end{eqnarray}
where $||.||$ is the induced norm. The Lanczos algorithm generalizes to the following systematic steps \cite{Lanczos:1950zz,vish123,Rabinovici:2020ryf}:

\begin{enumerate}

\item Set $b_{-1}\equiv 0$, $b_{0}\equiv 0$, and $|\mathcal{O}_{-1})= 0\,,$

\item Set $|\mathcal{O}_{0})= \frac{1}{||\mathcal{O}||}|\mathcal{O})\,,$

\item For $n \geq 1$: $|\mathcal{A}_{n})=\mathcal{L}|\mathcal{O}_{n-1}) - b_{n-1}|\mathcal{O}_{n-2})\,,$

\item with $b_{n}= ||\mathcal{A}_{n}||\,,$

\item If $b_{n}= 0$ stop; otherwise set $|\mathcal{O}_{n})=\frac{1}{b_{n}}|\mathcal{A}_{n})$ and go to step 3.

\end{enumerate}
The above algorithm constructs an orthonormal basis for $\mathcal{H}_{\mathcal{O}}$, $\Big\{ |\mathcal{O}_{n}) \Big\}^{K-1}_{n=0}$ known as the Krylov basis and a set of coefficients $\Big\{b_{n}\Big\}^{K-1}_{n=0}$ known as the Lanczos coefficients.
Thus, starting from $|\mathcal{O}_{0})$, each iteration of the Lanczos algorithm produces an element $|\mathcal{A}_{n})$ orthogonal to all previous Krylov basis elements $|\mathcal{O}_{m})$, with $m < n$. 
 $|\mathcal{A}_{n}) \neq 0$ for all $n < K$ as the orthogonal basis $\mathcal{K}$ has a rank $K$ or $K$ independent directions. At the next iterative level, $|\mathcal{A}_{K})$ is orthogonal to 
 $\Big\{|\mathcal{O}_{n})\Big\}^{K-1}_{n=0}$, which is already a complete orthonormal basis of $\mathcal{H}_{\mathcal{O}}$ and must therefore vanish. This explains the reason for $b_{K}=0$ in Step 5 and the termination of the iteration. Therefore, we conclude that the Lanczos algorithm must terminate once all the independent directions in $\mathcal{H}_{\mathcal{O}}$ are exhausted. \par
   One important aspect of writing the operator space $\mathcal{H}_{\mathcal{O}}$ in the Krylov basis $\mathcal{K}$ is that the Liouvillian $\mathcal{L}$ in the Krylov basis simplifies to a tridiagonal matrix
$(\mathcal{O}_{m}|\mathcal{L}|\mathcal{O}_{n})=\mathcal{L}_{mn}$. The entries of this tridiagonal matrix are given by the Lanczos coefficients. Alternatively,

\be{}
(\mathcal{L}_{mn})=\begin{bmatrix}
0 & b_1 & 0& \cdots& 0 &0& 0\\
b_1 & 0 & b_2 &\cdots &0 & 0 &0\\
0 & b_2 & 0 &\cdots &0 & 0 & 0\\
\vdots & \vdots & \vdots & \ddots &\vdots & \vdots & \vdots \\
0 & 0 &0 & \cdots& 0 &b_{K-2} &0 \\
0 & 0 &0 & \cdots &b_{K-2} &0 & b_{K-1}\\
0 & 0 &0 & \cdots & 0 & b_{K-1} & 0.
\end{bmatrix}
\ee

\par

 The Lanczos sequence $\{b_{n}\}$ behaves in a particular manner for chaotic and integrable systems. In both cases $\{b_{n}\}$ initially grows and then decreases to terminate for finite-dimensional systems. The nature of the $\{b_{n}\}$ sequence is fluctuating. Therefore, in \cite{Rabinovici:2021qqt} the authors have introduced the variance of Lanczos coefficients as a possible relation between the chaoticity of the system and it is given by,
\begin{eqnarray}
\label{variance_eqn}\sigma^{2}\equiv \text{Var}(x_{i})=\langle{x^2}\rangle
-\langle{x}\rangle^2,\,\,\,\,\, x_{i}\equiv \ln{\left(\frac{b_{2i-1}}{b_{2i}}\right)},
\end{eqnarray}
where $i$ runs from 1 to $K-1$ \footnote{Only the first half of the Lanczos sequence is used \cite{Rabinovici:2021qqt} for the statistics as the very small Lanczos coefficients at the end of the Lanczos sequence $\{b_{n}\}$ are numerically less reliable due to the amplification by the logarithm and the ratio in the definition Eq.\eqref{variance_eqn}.}, and $\langle \cdots \rangle$ represents the mean value. The variance $\sigma^{2}$ depends on the nature of the $b_{n}$ distribution. 

\par Before we end this section, we should keep in mind that, even after being a very powerful algorithm, the general Lanczos algorithm discussed above fails to construct an orthonormal basis often due to numerical instabilities. These numerical instabilities can occur due to the accumulation of errors at each level of the iteration. These instabilities can be removed using the reorthogonalisation technique discussed in the next section.

 \section{Reliability of Lanczos algorithm? } \label{Numerical}

One of the most significant consequences of using the Lanczos algorithm is that it is supposed to ensure the orthogonality of the Krylov basis obtained through each iteration in $n$. The algorithm is designed to stop when the Lanczos co-efficient $b_n$ hits zero. However, the Lanczos algorithm suffers major instabilities in numerical calculations \cite{parlett1998symmetric,articleSimon,10.2307/2006037,Rabinovici:2020ryf}. While computing the Krylov elements $|\mthA_n)$ in step 3 of the Lanczos algorithm described in the Sec.~\ref{KrySp}, we need the information of its previous two elements. Hence, the errors arising from the finite precision keep piling up, resulting in significant overlaps between the Krylov elements in each iteration. The more iterations, the faster the orthogonality is lost between the Krylov basis. This results in the Lanczos coefficients $b_n$ growing initially but never reaching zero. The $b_n$'s start to oscillate drastically around some average value after the initial growth. This unfaithful behaviour of the Lanczos coefficients was also reported in complex SYK \cite{Rabinovici:2020ryf}, and our analysis of the Bose-Hubbard model also went through similar errors.\par 
Nevertheless, there exist several ways to deal with the numerical errors piling up in the Lanczos algorithm, namely Full-orthogonalization (FO) and Partial re-orthogonalization \cite{Rabinovici:2020ryf, parlett1998symmetric,2020PhRvB.102s5419Y,PFO}. We found a useful resolution for our case, by implementing the Full-orthogonalization procedure.\footnote{For more details on the partial re-orthogonalization procedure, the readers are referred to \cite{Rabinovici:2020ryf}.} The key to implementing the FO algorithm is to execute the Gram-Schmidt at each iterative step $n$ twice and also use the information about all the Krylov elements (not only the previous two, as described in the previous section) to confirm the orthogonality between the Krylov basis (upto the error coming due to the machine precision). It is beneficial to explicitly mention the FO algorithm we have used successfully for the Bose-Hubbard Model. We start by computing the zeroth-operator $|\mthO_0)$ from the initial operator $|\mthO)$ using Eq.\eqref{defnorm} and step 2 of the Lanczos sequence. 
\label{stepRO}
\begin{enumerate}
\item Start with $|\mathcal{O}_{0})=\frac{1}{||\mathcal{O}||}|\mathcal{O})\,, $
\item Next Krylov element is $|\mthA_n)=\mathcal{L}|\mthO_{n-1})~~ \forall \;n\geq1\,,$ 
\item Perform re-orthogonalization using all of the previous Krylov elements by $|\mthA_n)\longrightarrow |\mthA_n)-\sum_{m=0}^{n-1}|\mthO_m)(\mthO_m|\mthA_n)$ (\textit{key difference from the usual Lanczos algorithm})\,,
\item Perform step 3 again (\textit{another key difference from the usual Lanczos algorithm})\,,
\item Now, define the Lanczos coefficient $b_n=\sqrt{(\mthA_n|\mthA_n)}\,,$
\item Next, set Krylov basis $|\mthO_n)=\frac{1}{b_n}|\mthA_n)$ and return to step 2\,,
\item Stop when $b_n=0\,.$
\end{enumerate}
 
  \section{Numerical results} \label{Results}

Now armed with the above discussion, we present our numerical analysis of operator growth for the Bose-Hubbard model, the Hamiltonian of which is given by Eq.\eqref{BHHam}. In particular, we study operator growth in Krylov space and the associated Lanczos coefficients respectively for a \textit{three-site and three-particle model}, ${\bf (M,N)=(3,3)}$, ${\bf (3,5)}$, and for a ${\bf (3,7)}$ model with different values of $\Lambda\equiv \frac{U}{J}$. We restrict ourselves to two ranges of values of $\Lambda$: $\Lambda\rightarrow{0}$, and $\Lambda=\text{finite}$ where we expect that the integrable and chaotic behaviour will appear following our previous discussion in Sec.~\ref{BHmod}.\par

First, we choose an appropriate hermitian operator for numerical computations. The operator that we choose is given by:
\begin{eqnarray}
\label{iniop}\mathcal{O}=\hat{n}_{j}+\hat{n}_{j+1}-i(a^{\dag}_{j}a_{j+1}-a^{\dag}_{j+1}a_{j}),\,\,\,\, j\in M
\end{eqnarray}
This choice of the operator has both the number operator at $j^{th}$ site $\hat{n}_{j}$ and the imaginary hopping term. \textit{One can, in general, also choose other operators, such as the number operator at a particular site $\hat{n}_{i}=a^{\dag}_{i}a_{i}$ or an imaginary hopping term $i(a^{\dag}_{j}a_{j+1}-a^{\dag}_{j+1}a_{j})$. We found that for all the choices mentioned above of the initial operator, the behaviour of the Lanczos coefficients $b_{n}$, and Krylov complexity $C_{K}$ typically remains the same \footnote{Although their numerical values will change for different choice of the initial operator.} and the conclusions drawn in this paper remain same.} {\color{black}One can further refer to Appendix~\eqref{A} where we have shown the numerical results for a different choice of initial operator \eqref{iniopdiff}.} The operator growth in the Krylov basis for the initial operator mentioned in Eq.\eqref{iniop} using the Lanczos algorithm is shown for the first two steps in the Appendix~\eqref{B}. Below, we discuss the numerical analysis done to find out the full Lanczos sequence for different system configurations. 
\begin{figure}[h]
\begin{subfigure}{0.45\linewidth}
 \includegraphics[width=\linewidth]{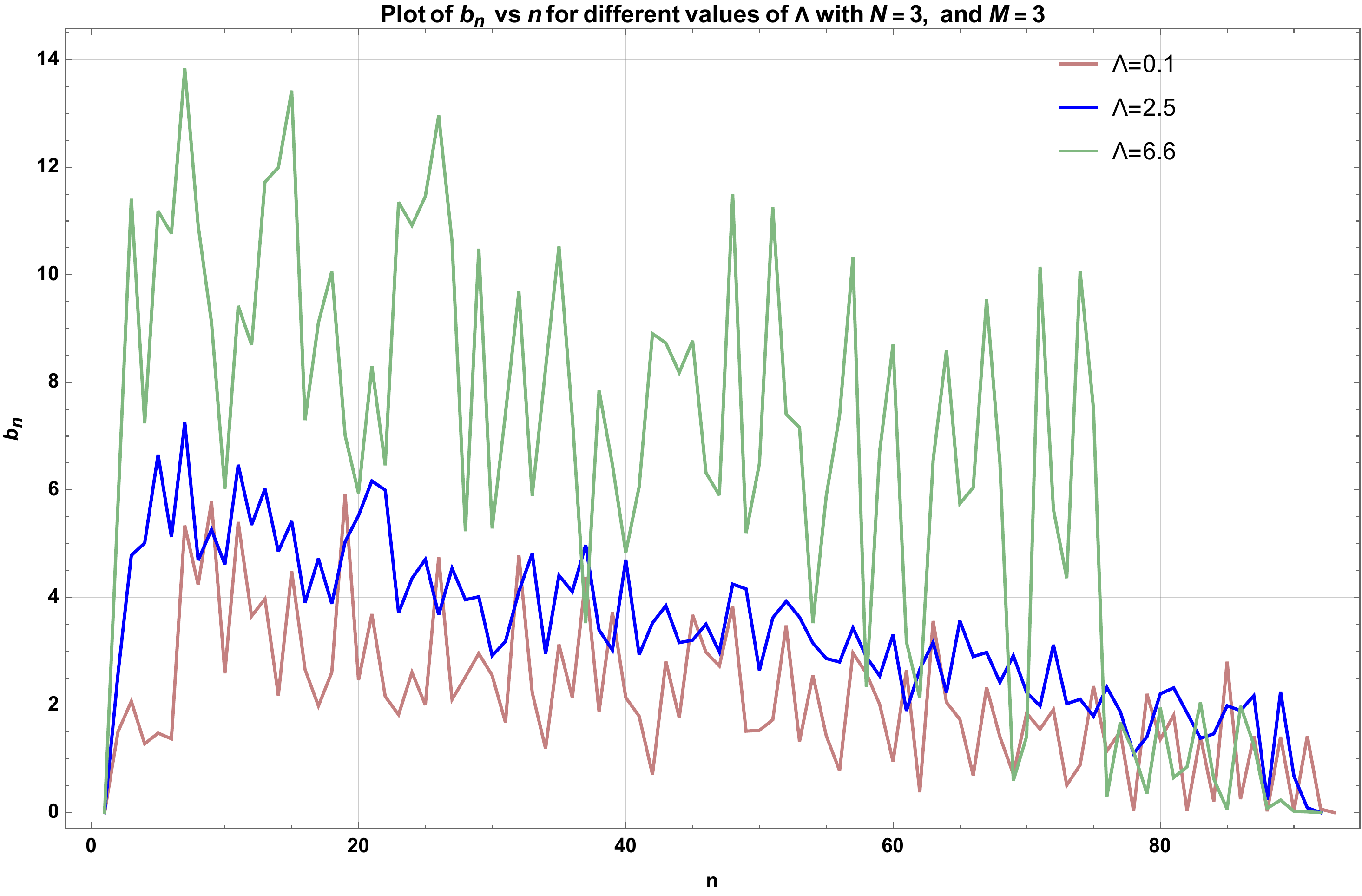}
 \caption{\sf Lanczos sequence for M=3, N=3. }
\label{fig:Lanczos_2.5}
\end{subfigure}
 \begin{subfigure}{0.45\linewidth}
 \includegraphics[width=\linewidth]{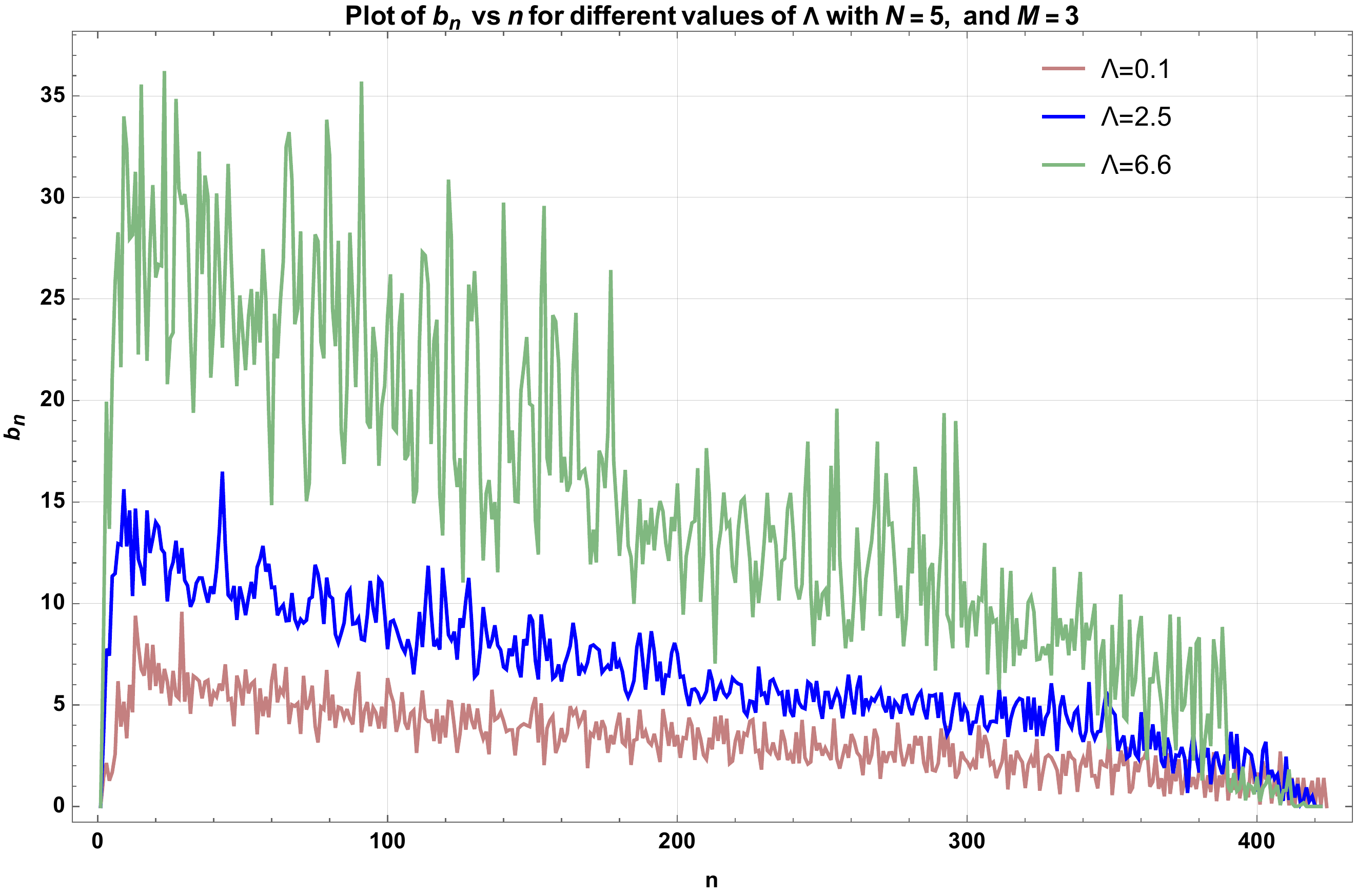}
 \caption{\sf Lanczos sequence for M=3, N=5.}
\label{fig:Lanczos_3,5}
\end{subfigure}
\hfill
\vspace{0.5cm}
\begin{subfigure}[b]{0.45\linewidth}
\includegraphics[width=\linewidth]{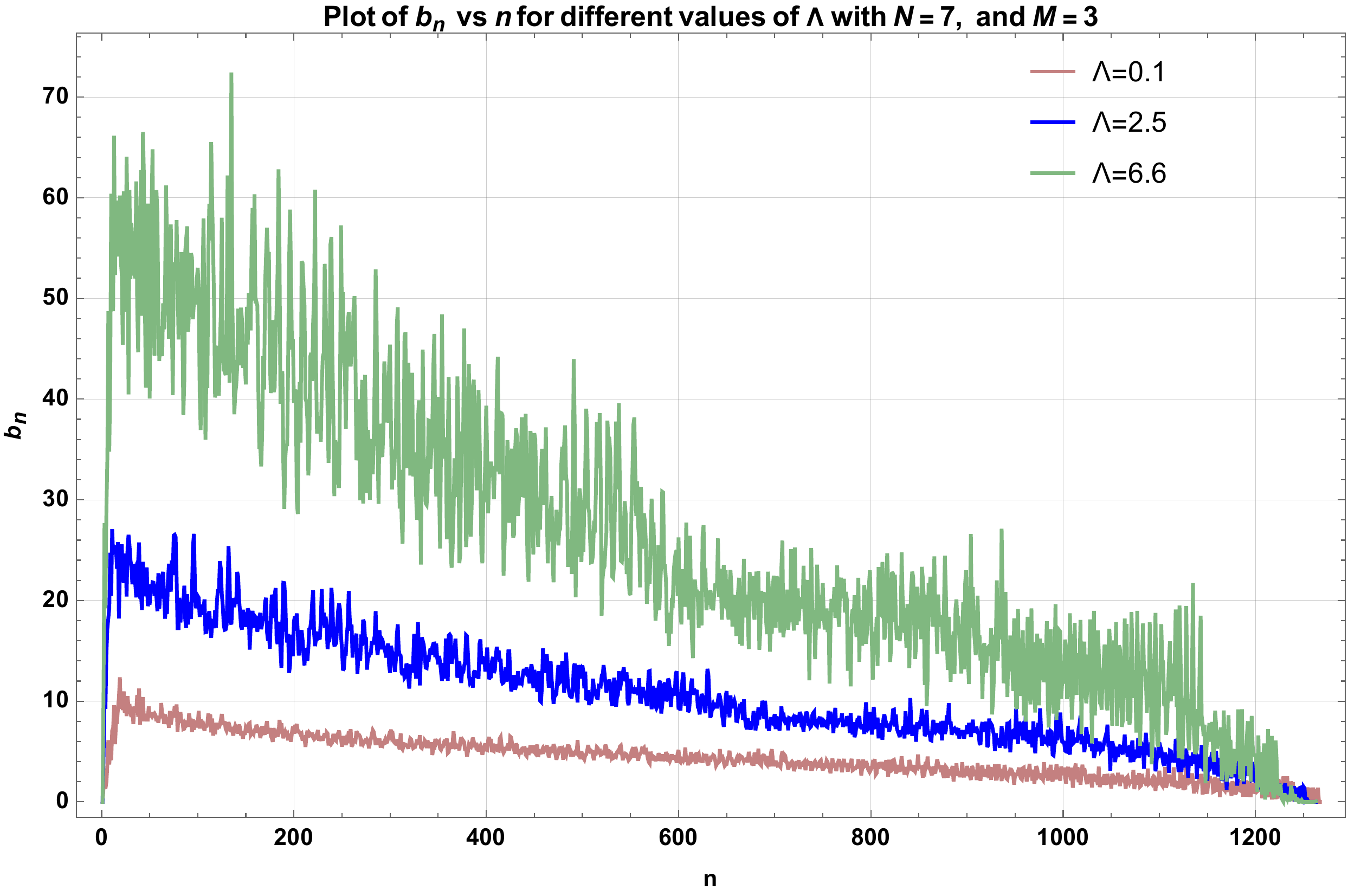}
\caption{\sf Lanczos sequence for M=3, N=7 }
\label{fig:fig:Lanczos_3,7}
\end{subfigure}
\begin{subfigure}[b]{0.45\linewidth}
\includegraphics[width=\linewidth]{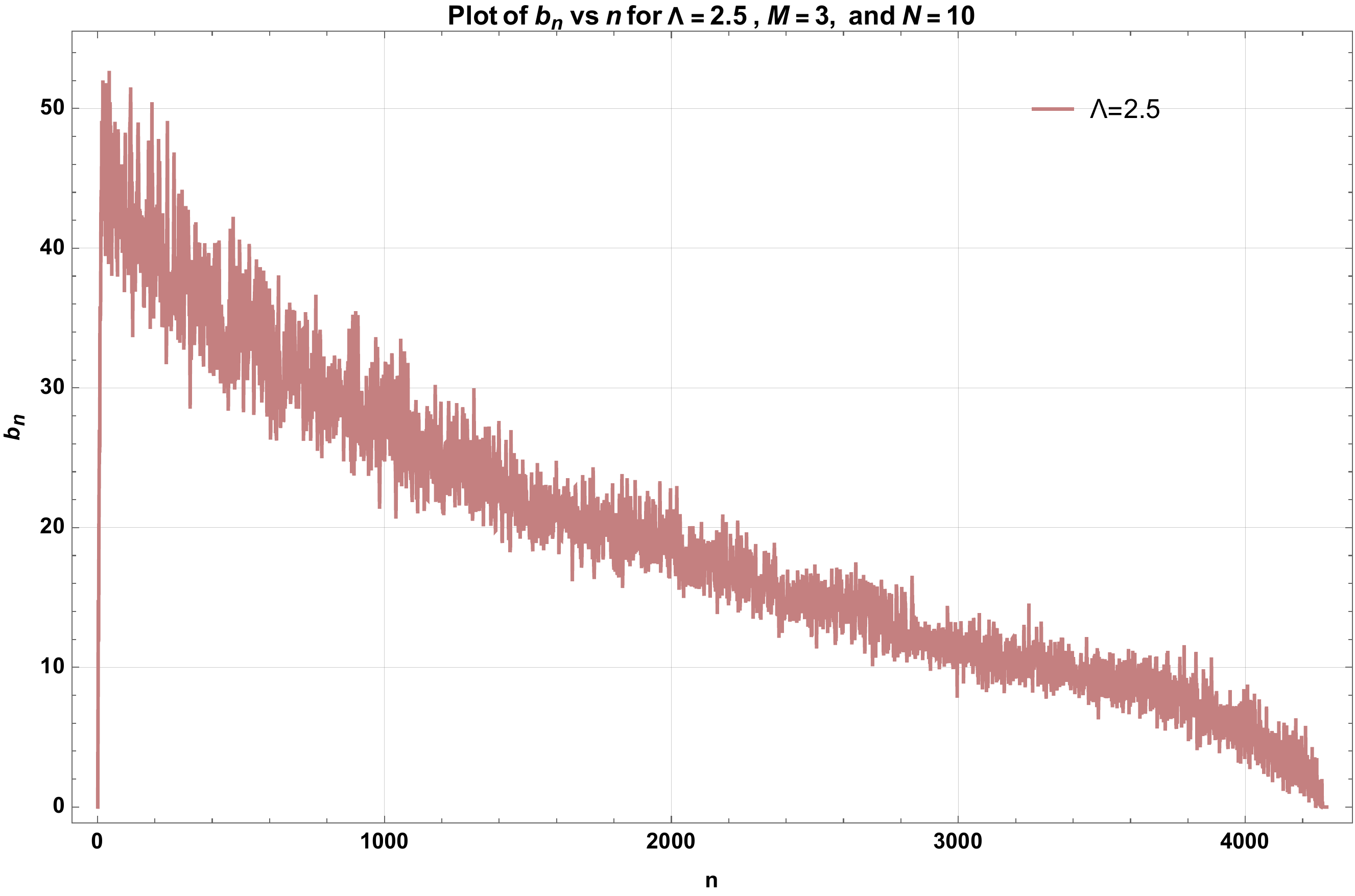}
\caption{\sf Lanczos sequence for M=3, N=10 }
\label{fig:Lanczos_3,10}
\end{subfigure}
\caption{\sf Lanczos sequence for different configurations of M, and N. Fig.(\ref{fig:Lanczos_2.5}) shows the Lanczos sequence for N=3, and M=3, Fig.(\ref{fig:Lanczos_3,5}) for  N=5, and M=3, and Fig.(\ref{fig:fig:Lanczos_3,7}) for N=7, and M=3 respectively for the values of $\Lambda=0.1, 2.5, 6.6$. Fig.(\ref{fig:Lanczos_3,10}) plots the Lanczos coefficients $b_{n}$ for $\Lambda=2.5$, N=10, and M=3. For all the cases, the Lanczos coefficients $b_{n}$ initially grows linearly with n up to $n\sim \ln{D}$, and then slowly decrease to zero with a slope of order $\sim -\frac{1}{D^{2}}\sim -\frac{1}{K}$. The Lanczos sequence $\{b_{n}\}$ smoothens as we increase the system size.}
\end{figure}

We restrict our analysis to the a three-site Bose-Hubbard model with particle numbers $N=3,5,7$. We state our numerical results for the three-site Bose-Hubbard model with an initial operator mentioned in Eq.\eqref{iniop1}, and for different values of interaction strength parameter $\Lambda$ as following:
\begin{itemize}

\item {\color{black}For $\Lambda=0$, the Lanczos iteration stops at the fifth iterative order  with $b_{5}=0$. So, the dimension $K=4$ of Krylov basis is much smaller than the dimension of the Hilbert space of operators $D^{2}=100$ and hence, confirms the bound  mentioned in Eq.(\ref{Kbasisbound}). For
smaller values of $\Lambda\sim 10^{-7}-10^{-3}$, the Lanczos iteration stops at the fifth iterative order }\footnote{\color{black}Beyond this order the Lanczos coefficients $b_{n}$ fails to follow the linear growth-plateau behavior. More precisely, $b_{n}$ becomes highly irregular.} {\color{black}with $b_{5}\sim 10^{-6}.$ This kind of behavior is expected  as the Bose-Hubbard model is learned to be integrable \cite{Nakerst:2022prc} for $\Lambda\rightarrow 0$ as $\Lambda=0$ or $U=0$ is just a Gaussian model and hence this is consistent.\footnote{\textcolor{black}{In \cite{Nakerst:2022prc}, it was shown that for very small (but non-vanishing) $\Lambda$ all the Lyapunov's are zero. Our result is consisting with this fact.}}}

\item Similarly, for large $\Lambda\sim 10^{15}$, the Lanczos iteration \footnote{\color{black} At larger values of $\Lambda\sim 10^{2}$, the Lanczos coefficient $b_{n}$ becomes highly irregular and does not exhibit the linear growth-plateu behaviour similar to smaller values of $\Lambda$} stops at a very small value of $K$, which is in agreement with the integrable nature of the Bose-Hubbard model at $\Lambda\rightarrow \infty$ \cite{Nakerst:2022prc} as in this limit, the only possible Lyaponuv exponent is found to be exactly zero.

\item The Lanczos iteration saturates the upper bound mentioned in Eq.(\ref{Kbasisbound}) for finite values of $\Lambda$. The plots of $b_{n}$ with $n$ show a phase of initial growth up to $n\sim \ln{D}$. The initial phase of growth is then followed by a regime of steady decrease towards zero with a generally constant negative slope of the order $\sim -10^{-2}\sim K^{-1}$. This is shown in the Fig.~(\ref{fig:Lanczos_2.5}), (\ref{fig:Lanczos_3,5}) and (\ref{fig:fig:Lanczos_3,7}) for different values of $(M,N)$. 
\end{itemize}
\vspace{-0.25cm}
These observations suggest that the operator growth via the nature of the Lanczos coefficients can distinguish the system as being chaotic or non-chaotic (integrable), providing further evidence to the earlier claims \cite{Parker:2018yvk,Rabinovici:2022beu,Rabinovici:2021qqt}. We summarize our conclusions regarding the nature of the Lanczos coefficients in the Table~\ref{table:b}.
\begin{table}[h]
\centering
\begin{tabular}{|c|c|}
\hline
 $\mathbf{n}$            & \textbf{Nature of Lanczos coefficients} $\mathbf{b_{n}}$ \\
\hline
$1<<n< \ln{D}$ & $b_{n}$ grows linearly with n: $b_{n}\sim c\,n$  \\
\hline
$\ln{D}<<n<D^{2}$ & a steady fall-off of $b_{n}$ with n having slope $\sim \frac{1}{K}$  \\
\hline
$n \sim D^{2}$   & $b_{n}$ falls off to zero  \\
\hline
\end{tabular}
\caption{The table summarises different behaviors of Lanczos coefficients $b_{n}$ with n for a chaotic system having the dimension of Krylov basis $K$.}
\label{table:b}
\end{table}

\subsection*{Variance of Lanczos coefficients}
\begin{figure}[htb!]
\begin{center}
 \includegraphics[width=0.7\textwidth]{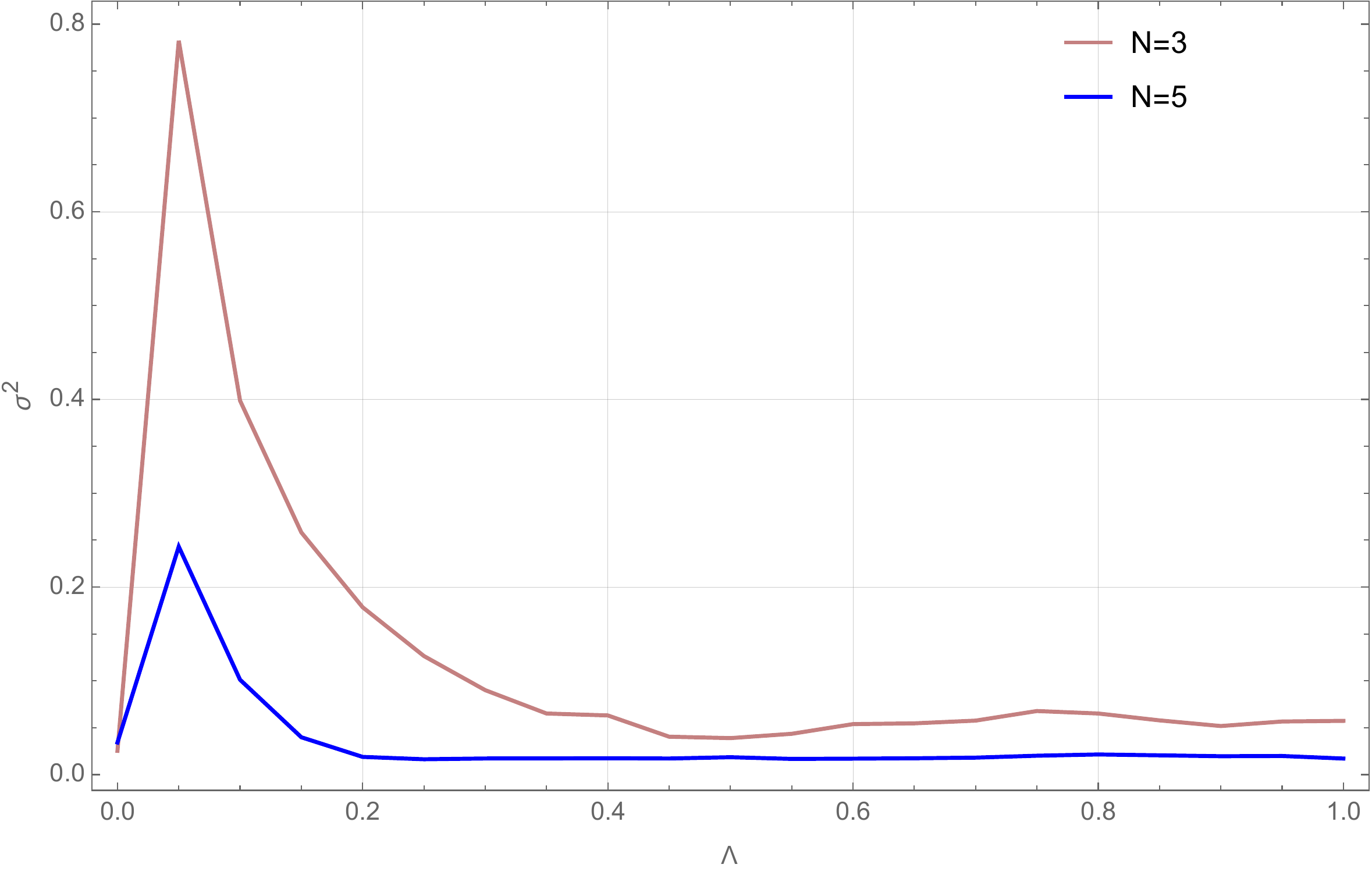}
 \caption{\sf The above figure plots the variance $\sigma^2$ as a function of the interaction parameter $\Lambda$ for $(N,M)= (3,3),(3,5)$. It can be seen from the plot that as $\Lambda$ becomes finite starting from $\Lambda=0$, the variance decreases and becomes close to zero for $\Lambda=1$. This behaviour of the variance points towards the transition from integrable to chaotic. }
\label{fig:variance_plot}
\end{center}
\end{figure}

Before we end the section, \textcolor{black} {we numerically study the variance \footnote{In \cite{Balasubramanian:2022dnj}, the authors have studied different Lanczos coefficients $b_n$ for Random Matrix Theories. They have shown that for different ensembles (GOE, GUE, and GSE) and generic initial state, the Lanczos coefficients $b_{n}$ are independent random variables and the nature of their variance is of the same form as shown in  Fig.~\ref{fig:variance_plot}. } of ${b_{n}}$ due to their fluctuating behaviour, for the integrable and chaotic regimes \footnote{A similar analysis has been done in \cite{Hashimoto:2023swv} for stadium billiards as a function of the shape of the billiards.} 
as a function of the parameter $\Lambda$}.
\begin{itemize}
    \item In Fig.~\ref{fig:variance_plot}, we have shown the nature of the variance $\sigma^2$ as a function of the interaction parameter $\Lambda$. 
    We notice that as the value of the parameter $\Lambda$ increases from 0 to 1, the variance $\sigma^2$ becomes closer to zero. \textcolor{black}{The above behavior of the variance is typically true for chaotic systems \cite{Hashimoto:2023swv,Rabinovici:2021qqt} and hence we conclude that the system is integrable for $\Lambda=0$ and becomes more and more chaotic as $\Lambda$ approaches the value 1.}

    \item \textcolor{black}{Also, as we increase the system size, the Lanczos sequence smoothens out for the chaotic system, bringing the variance closer to zero and causing a more smooth transition from the integrable regime, $\Lambda=0$ to the chaotic one  $\Lambda> 0$.}

    \item \textcolor{black}{One can notice a sharp fall-off in $\sigma^2$ values as the value of the $\Lambda$ increases from \textit{zero}. It marks the transistion of the system from a integrable to a chaotic one as was expected from our previous analysis of the Lanczos coefficients. }

    \item \textcolor{black}{Also, one can find that a similar sharp peak can be obtained at $\Lambda \sim \mathcal{O}(10^2)$ which marks the transition from chaotic to an integrable system.} \footnote{\color{black}{We notice that the profile of $b_{n}$ at larger values than $\sim \mathcal{O}(10^2)$ becomes highly irregular with no ramp-plateau behavior. Therefore, we can safely say that the system has indeed transitioned to an integrable system from a chaotic one. }}
     \item \textcolor{black}{Also, in Fig.~\ref{fig:variance_plot} one can notice ``kinks'' near $\Lambda \sim 0.1$, $\Lambda \sim 0.15$ etc. These kinks are due to the unsmooth behavior of Lanczos coefficients for small system size. As we increase the system size and $b_{n}$s become smooth, these kinks disappear. One can immediately get a hint of this from  the $N=5$ plot where some of the kinks disappear.}
\end{itemize}
\textcolor{black}{We have listed our findings from the numerical analysis of $b_n$ pointwise above where we have analysed three-site Bose-Hubbard model for different system configurations. Note that the disordered average of $b_{n}$ has been considered in \cite{Rabinovici:2020ryf} for the SYK$_{4}$ model and in \cite{Trigueros:2021rwj} for a many-body localization (MBL) systems, e.g., one-dimensional (1D) spin chain with $L$ sites and open boundary conditions, where in both cases values for the interaction strengths are drawn randomly from a certain random distribution and then $\{b_{n}\}$ are averaged over all the random realizations. The interaction parameter $\Lambda$ in the Bose-Hubbard model does not follow from a random distribution. We do our analysis for particular values of $\Lambda$ to characterize the chaotic and non-chaotic (integrable) behaviour.} 

\section{Krylov Complexity} \label{Results1}
\textcolor{black}{Now we investigate the \textit{Krylov complexity}, often termed as \textit{K-complexity} that was introduced by Parker et.al \cite{Parker:2018yvk} as a useful diagnostic of quantum chaos in thermodynamic limit which measures an operator's complexification over
the orthonormal Krylov basis. \cite{Parker:2018yvk}}. \textcolor{black}{Also motivated by our study of the variance of Lanczos coefficients in the previous section, we will mainly focus on the finite $\Lambda$ regime where the model is expected to show chaotic behaviour.} It is obtained by first constructing the Krylov basis using the Lanczos algorithm given by \eqref{stepRO} and then using the Lanczos coefficients $b_{n}$ to reduce the analysis of the time-evolution of an operator $\mathcal{O}$ into the solution of a  \textcolor{black} {differential recurrence equation \cite{Parker:2018yvk}. The definition of Krylov complexity is outlined below.}\par

The time-evolved operator can be expanded in the Krylov basis as:

\be{}
\label{time_evol}|\mthO(t))=e^{it\mathcal{L}}|\mthO)=\sum_{n=0}^{K-1}\phi_n(t)|\mthO_n),
\ee
where $\phi_n(t)$ can be thought of as single-particle wavefunctions distributed over the Krylov basis, which via the Heisenberg equation
$\frac{d\mathcal{O}}{dt}=i[H,\mathcal{O}]$ satisfies the following differential recurrence equation,
\be{}
\label{recurrence_eqn}-i\dot{\phi}_{n}(t)=\sum_{m=0}^{K-1}L_{mn}\phi_m(t)=b_{n+1}\phi_{n+1}(t)+b_{n}\phi_{n-1}(t).
\ee
To solve this equation we need an initial conditions which is: $$\phi_{n}(t=0)=\delta_{n0},$$  For a normalised operator $\mathcal{O}(0)=\mathcal{O}_{0}$, and $\phi_{-1}(t)\equiv 0$ to make Eq.~\eqref{recurrence_eqn} consistent with Eq.~\eqref{time_evol} we need to impose this initial condition. Furthermore, the boundary conditions $b_0=b_K=0$ are fixed to ensure the finiteness of this operator-evolution problem. One can also infer from the equation of motion  mentioned Eq.~\eqref{recurrence_eqn} of the wavefunctions $\phi_n(t)$
as a single-particle hopping problem on a semi-infinite chain called the {\textit{Krylov chain}}, with the hopping amplitudes given by the Lanczos coefficients $b_{n}$ and $\phi_n(t)$ are the wave-packets \textcolor{black}{traveling on the chain \cite{Parker:2018yvk}}.
Since the initial operator is normalized at the first step of the Lanczos algorithm  mentioned in Sec.~\eqref{stepRO}, the notion of unitarity suggests that the wavefunction $\phi_n(t)$ is normalized at all times: $$\sum_{n=0}^{K-1}|\phi_n(t)|^2=1.$$

Then Krylov complexity or K-complexity is defined as the time-dependent average position of the distribution $\phi_n(t)$ over the  \textcolor{black}{ordered Krylov basis \cite{Parker:2018yvk}}:
\be{}
\mathcal{C}_K(t)=\sum_{n=0}^{K-1}n|\phi_n(t)|^2.
\ee

We discuss the behavior of the K-complexity emerging from the numerical analysis in the following points:

\begin{figure}[h!]
\begin{center}
 \includegraphics[width=0.47\textwidth]{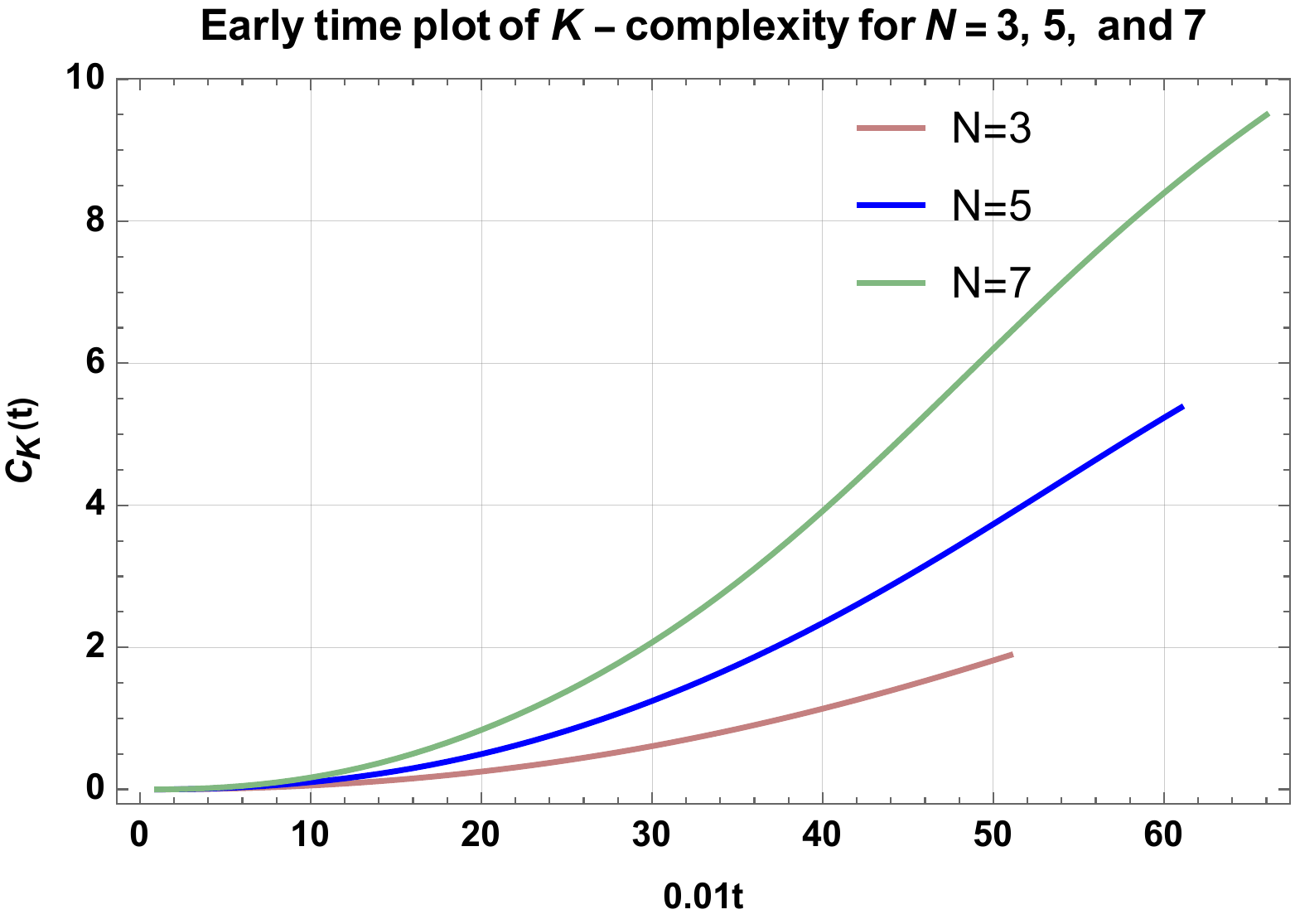}
 \caption{\sf Early time plot ({\it scrambling}) of Krylov complexity $C_{K}(t)$ for $\Lambda=2.5$ with time t. The magenta, blue, and green line represents plot of  $C_{K}(t)$ for $\{N=3, M=3, t_{s}=0.36\}$, $\{N=5, M=3, t_{s}=0.48\}$, $\{N=7, M=3, t_{s}=0.55\}$ respectively.}
\label{fig:ETCom2533}
\end{center}
\end{figure}

\begin{itemize}
\item We observe an exponential growth in K-complexity $\mathcal{C}_K(t)$ at very early times until $t_{s}=\log{(\log{D})}$. This is known as the scrambling time. The value of scrambling time $t_{s}$ changes as the system configuration changes which is shown in Fig.~\ref{fig:ETCom2533}.

\begin{figure}[h!]
\begin{subfigure}{0.47\linewidth}
\centering
 \includegraphics[width=\linewidth]{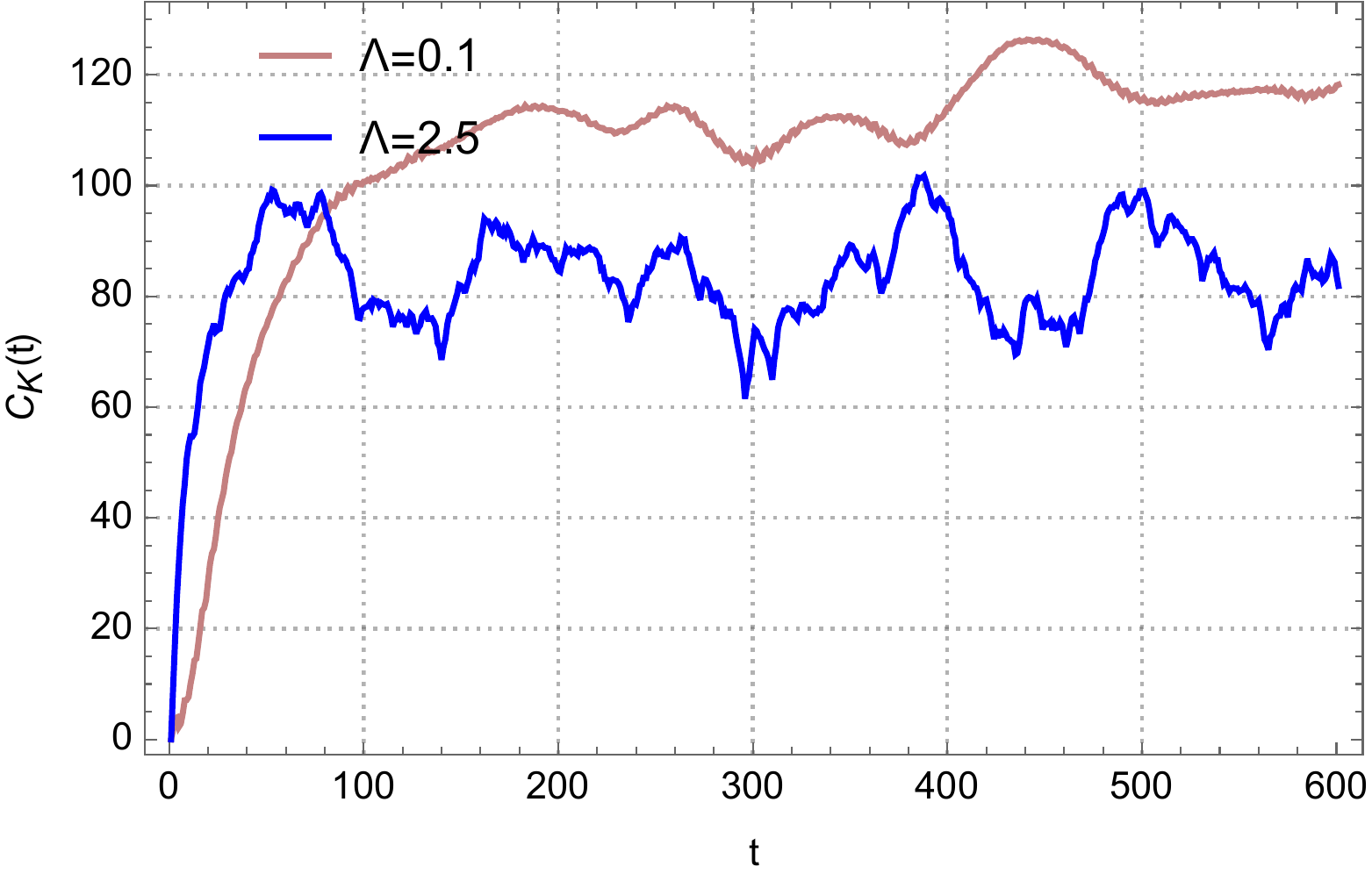}
 \caption{}
 \end{subfigure}
 \begin{subfigure}{0.47\linewidth}
  \includegraphics[width=\linewidth]{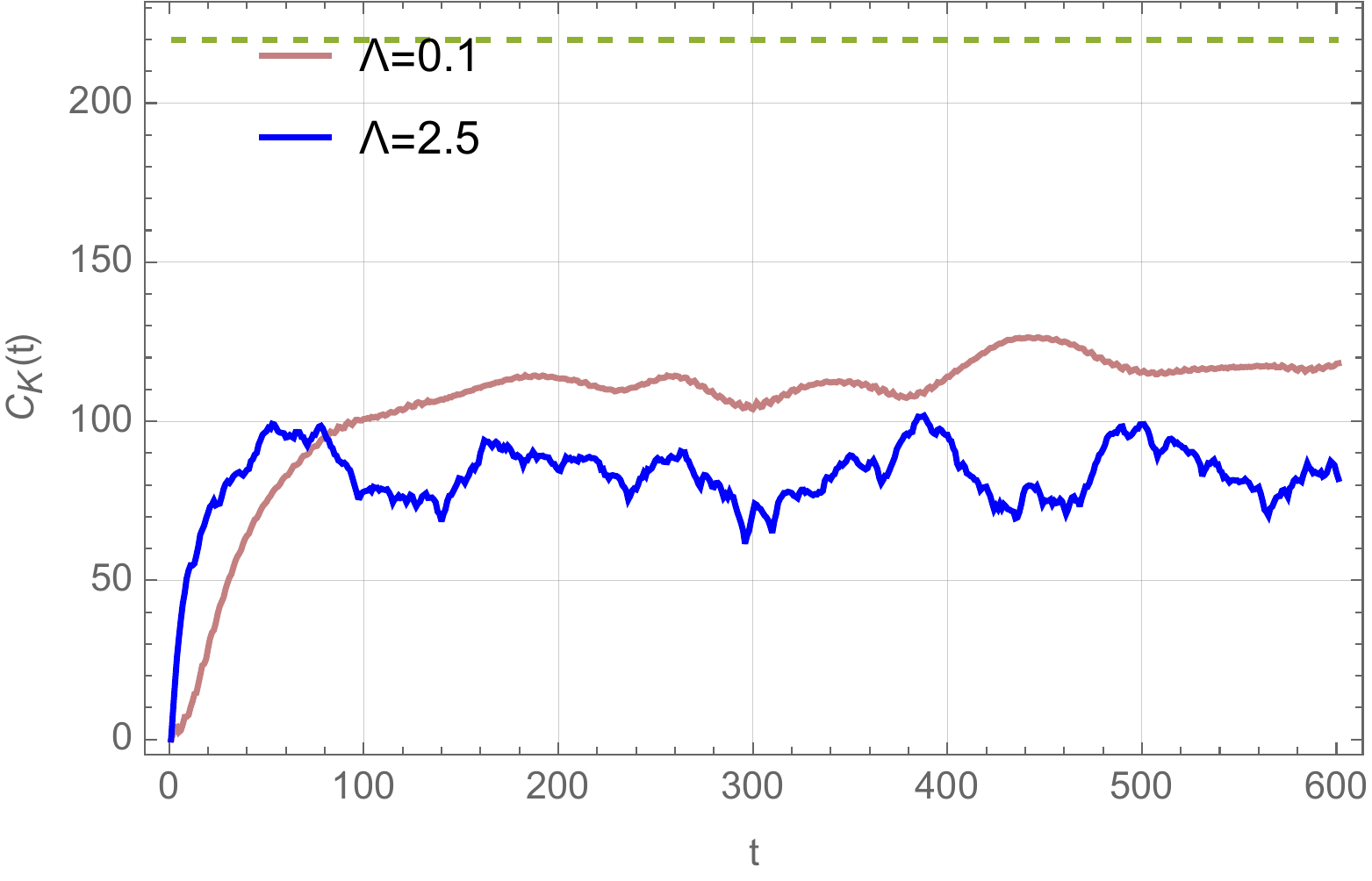}
  \caption{}
   \end{subfigure}
 \caption{\sf Plot of K-complexity at all time scales for $\Lambda=0.1, 2.5$ with $N=5$, and $M=3$. The linear growth and the saturation can be observed at times of $\mathcal{O}(D)$, and at $\mathcal{O}(D^{2})$ respectively. For $N=5$, and $M=3$, $D=441$ and $K=421$. The saturation value of $\mathcal{C}_{k}(t)$ is $\frac{K}{2}=220$ is shown by the dashed line in the second figure.}
\label{fig:FCom2533}
\end{figure}

\item After the scrambling time $t_{s}$, the K-complexity $\mathcal{C}_K(t)$ features a linear increase in time until times of order $t\sim D$ and finally at very late times of order $t\sim D^2$ it saturates (with fluctuations around that saturation value as shown in Fig.~\ref{fig:FCom2533}) typically at the values\footnote{This is in agreement with \cite{Rabinovici:2021qqt, Rabinovici:2022beu}. In general, the complexity saturation values are below $\frac{K}{2}$. For more details on the saturation value, the readers are referred to \cite{Rabinovici:2022beu}.} of order~$\frac{K}{2}$. This phenomenon can be understood physically as a uniform distribution of operator in Krylov basis. The behavior of the K-complexity at all time scales is shown in Fig.~\ref{fig:FCom2533} for different values interaction parameter $\Lambda$.
\end{itemize}
The above results are summarised in Table~\ref{table:c}. The characteristics of K-complexity described above clearly agree with the results of \cite{Barbon:2019wsy} where the profile of K-complexity for generic operators in a finite-size system was first shown \footnote{It was observed in \cite{Bhattacharyya:2020art}, that the system consists of $N$ coupled oscillators in the inverted regime for large $N$, three different time scales of the same order as found here for Nielsen's circuit complexity \cite{NL1,NL2,NL2} emerge. However, the behaviour of the circuit complexity in those regimes, particularly in the first two, is different from what we have obtained here via Krylov complexity.}. {\color{black} One can refer to Appendix.\eqref{A2} where numerical fit 
is performed to analyse the behaviour of K-complexity at different time scales}.
\begin{table}[!ht]
\centering
\begin{tabular}{|c|c|}
\hline
 $\mathbf{t}$            & \textbf{Nature of Krylov complexity} $\mathbf{C_{K}(t)}$ \\
\hline
$0\leq t\leq \log{(\ln{D})}$ & $C_{K}(t)$ grows exponentially in time t : $C_{K}(t)\sim \exp{c\,t}$  \\
\hline
$t\sim D$ & linear growth of $\mathcal{C}_{K}(t)$ with t  \\
\hline
$t \sim D^{2}$   & $C_{K}(t)$ saturates \\
\hline
\end{tabular}
\caption{The table distinguishes different behaviors of Krylov complexity $C_{K}(t)$ with t  for a chaotic system.}
\label{table:c}
\end{table}
\section{Discussion} \label{discussion}
In this article, we considered the Bose-Hubbard model as a quantum many-body model that is neither integrable nor strongly chaotic. It also has a quartic term in the Hamiltonian, providing a platform to study   Krylov complexity beyond the quadratic discrete bosonic quantum many-body system. The ultra-cold atom Hamiltonian is directly related to the Bose-Hubbard Hamiltonian. We investigate a considerably more basic form of the model with a chemical potential $\mu=0$ and, thus, does not exhibit any phase transition.\par

We start by analyzing the operator growth in Krylov basis in a three-site $(M=3)$ Bose-Hubbard model for different parameter $\Lambda$ values. Furthermore, we perform a numerical analysis to explore operator growth in the Bose-Hubbard model. We study cases with different particle sizes, $N$. \textcolor{black}{Although we have shown the results for a specific initial operator, overall features remain the same even if we consider different initial operators.} \textit{An important observation of our work is that the standard method of performing the Lanczos algorithm fails for this model due to the piling of errors at each level of the iteration. We remove this difficulty by performing the Full orthogonalization (FO) method to find the appropriate Krylov basis and the Lanczos coefficients}.\par
One of our crucial findings was that the dimension of the Krylov basis $(K)$ is much smaller than the dimension of the Hilbert space for the operators when $\Lambda \rightarrow 0$ or $\Lambda \rightarrow \infty$. This confirms the fact that the three-site Bose-Hubbard model becomes integrable at the asymptotic limits of $\Lambda$. The Lanczos coefficient $b_n$ displays ramp and plateau behaviour for finite values of $\Lambda$. This indicates that for finite values of $\Lambda$, the three-site Bose-Hubbard model becomes chaotic, and the dimension of the Krylov basis saturates the upper bound as mentioned in Eq.~\eqref{Kbasisbound}. \textcolor{black}{Also, in addition to the study of the behaviour of Lanczos coefficients $b_{n}$, we also study the variance $\sigma^2$ of the Lanczos sequence as a function of the interaction parameter $\Lambda$  which arises naturally as $\{b_{n}\}$ are fluctuating in nature. We observed that the variance follows the typical nature of being close to zero \cite{Hashimoto:2023swv} as we approach from an integrable system to a chaotic one. It will be interesting to figure out in future how we can extract the information about spectral statistics, which encodes the information of the quantum chaos from the statistics of $b_n$ perhaps following the study of \cite{Erdmenger:2023shk, Hashimoto:2023swv}.} \par

Furthermore, we investigate the Krylov complexity for finite $\Lambda$. We have investigated the behaviour of K-complexity at three different time scales. We discovered that the K-complexity exhibits exponential increase at initial time scales, typically of $\mathcal{O}(\log(\ln{D}))$, followed by a linear growth over time. K-complexity saturates in time at very late times, of the order of the dimension of the operator Hilbert space. This demonstrates that the three-site Bose-Hubbard model becomes chaotic for finite $\Lambda$. \textcolor{black}{The emergence of these three distinct time scales matches with the general claim made in the literature, e.g. \cite{Barbon:2019wsy}}. \par
There are various future directions that will be interesting to pursue. {\color{black}One can study the Liouvillian spectrum and operator matrix elements in energy eigenbasis to find universal chaotic behaviour of the system from ETH approach\cite{progress}}. Throughout our studies, we have set the chemical potential to zero.  However, the system can undergo quantum phase transitions in the presence of non-vanishing chemical potential. It will be interesting to perform our analysis for this case and whether we can comment on the phase transition using Krylov complexity. This will provide a possibility of comparing with the results coming from the circuit complexity analysis for the Bose-Hubbard model as well as connecting with results from holography \cite{Sood:2021cuz,Sood:2022lfx,Huang:2021xjl}. Also, it will be interesting to extend our computation for finite temperature case and compare with the results of \cite{Avdoshkin:2019trj,Avdoshkin:2022xuw,Camargo:2022rnt} whether we get a staggered-like behaviour for Lanczos coefficients. Recently emergence is scar-like states which evade thermalization has been demonstrated in Bose-Hubbard type models \cite{2020PhRvL.124p0604Z,2022arXiv221212046H,Su:2022glk}. It will be interesting to study the (spread) complexity of such states along the line of \cite{Nandy:2023brt}. Furthermore, we like the continuous limit of this model study the Krylov complexity and hope to connect with the recent studies of it for field theories \cite{Avdoshkin:2019trj,Camargo:2022rnt,Dymarsky:2021bjq}. We hope to address some of these issues in the near future and report on them. Last but not least, as mentioned earlier, the Bose-Hubbard model emerges from various physical scenarios involving ultra-cold atom gas in periodic potential (e.g. dimerized magnets \cite{2012arXiv1210.1270L}, interaction-induced tunnelling \cite{2012PhRvL.108k5301S}) and can potentially be used as a quantum-simulator \cite{Yang:2020yer}, we hope that our study of operator growth and the associated complexity for this system might shed important light in this context and initiate further studies.

\section*{Acknowledgements}
D.G would like to thank the Physics department at the Indian Insititute of Technology Gandhinagar for their hospitality and Tanmoy Sengupta, Sumit Shaw, Ajit C. Balaram, Sashikanta Mohapatra, and Sayan Mukherjee for their useful suggestions and discussions. A.B like to thank the FISPAC Research Group, Department of Physics, University of Murcia, especially, Jose J. Fernández-Melgarejo for hospitality and useful discussion based on the seminar given on this paper. A.B and P.N is supported by Relevant Research Project grant (202011BRE03RP06633-BRNS) by the Board Of Research In Nuclear Sciences (BRNS), Department of Atomic Energy (DAE), India. A.B is supported by Mathematical Research Impact Centric Support Grant (MTR/2021/000490) by the Department of Science and Technology Science and Engineering Research Board (India). A.B and P.N thank the speakers and participants of the workshop ``Quantum Information in QFT and AdS/CFT-III" organized at IIT Hyderabad between 16-18th September, 2022 and funded by SERB through a Seminar Symposia (SSY) grant (SSY/2022/000446) and  ``Quantum Information Theory in Quantum Field Theory and Cosmology" between 4-9th June, 2023 hosted by Banff International Research Centre at Canada for useful discussions. A.B. also acknowledge the associateship program of the Indian Academy of Science, Bengaluru.
\newpage
\appendix 
\section{An alternate choice of initial operator}\label{A}
\begin{figure}[h!]
\begin{subfigure}{0.47\linewidth}
\centering
 \includegraphics[width=\linewidth]{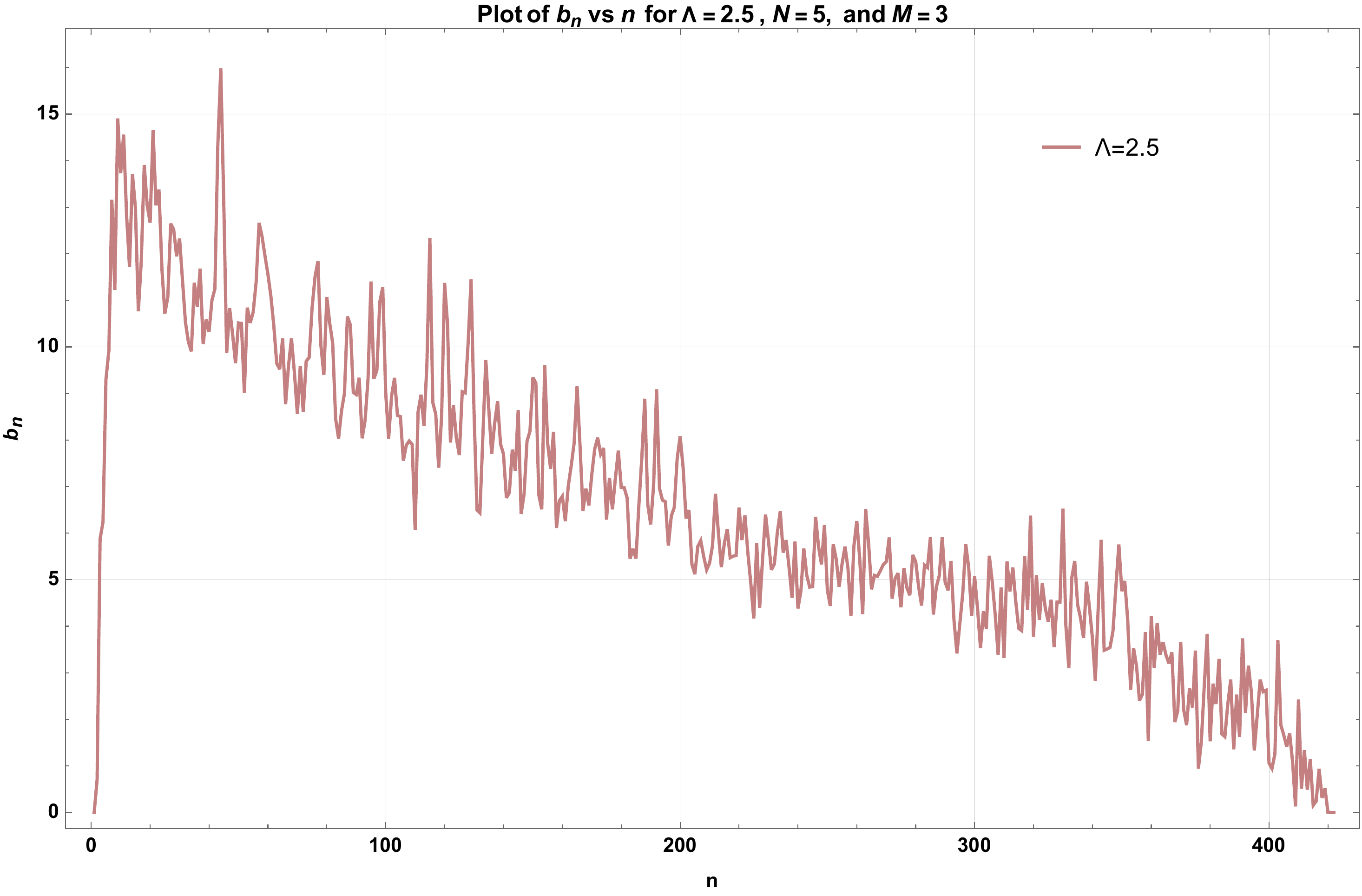}
 \caption{}
 \label{fig:diffopLan_2.5}
 \end{subfigure}
 \begin{subfigure}{0.47\linewidth}
  \includegraphics[width=\linewidth]{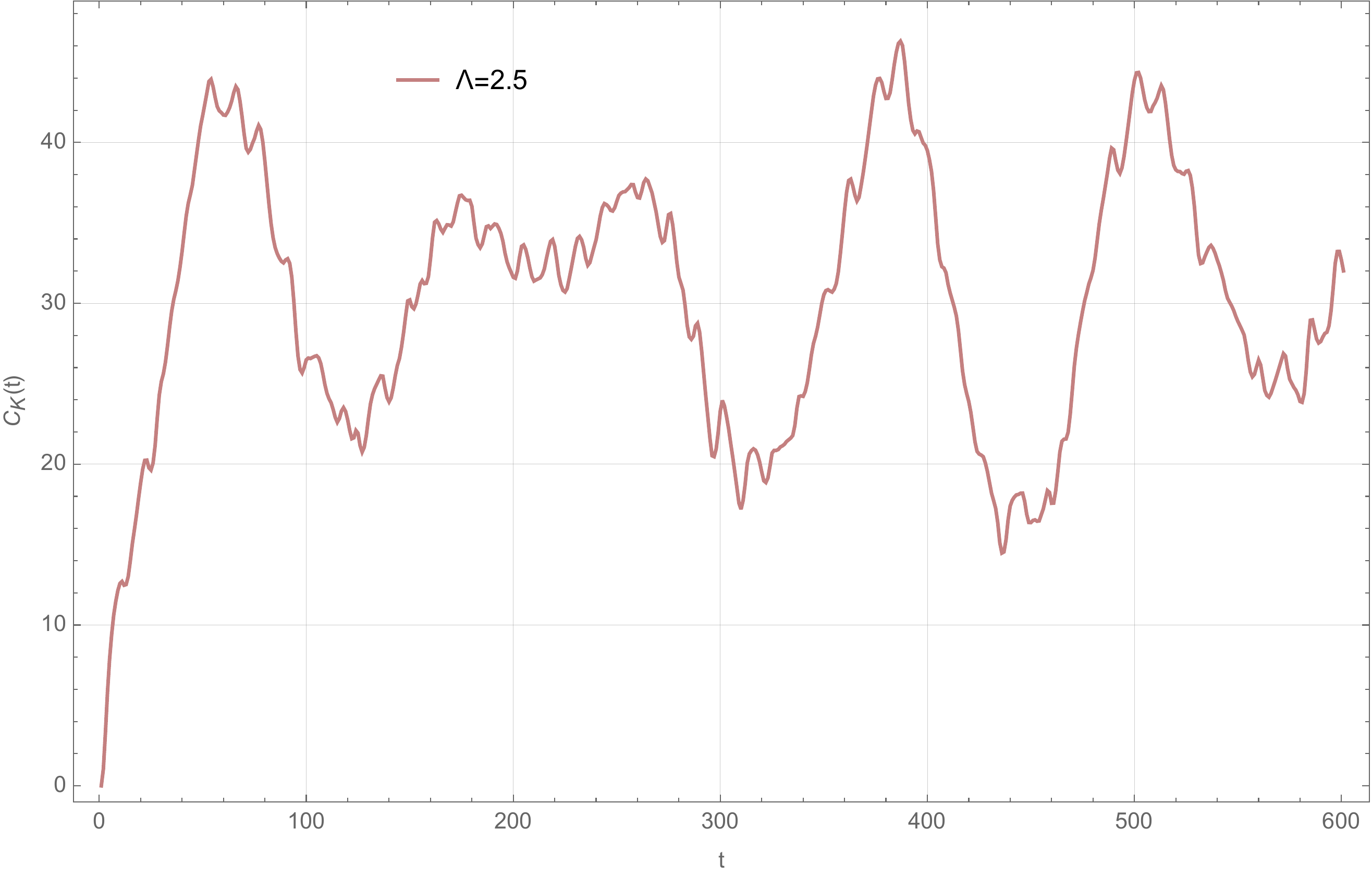}
  \caption{}
  \label{fig:diffopCom_2.5}
   \end{subfigure}
 \caption{\sf Lanczos sequence Fig.(\ref{fig:diffopLan_2.5}) and K-complexity Fig.(\ref{fig:diffopCom_2.5}) for a different choice of initial operator \eqref{iniopdiff}. Fig.(\ref{fig:diffopLan_2.5}) shows the Lanczos sequence for $N=5$, and $M=3$. One can easily see and relate to the previous results and conclude that the Lanczos coefficients $b_{n}$ initially grow linearly with n up to $n\sim \ln{D}$, and then slowly decrease to zero with a slope of order $\sim -\frac{1}{D^{2}}\sim -\frac{1}{K}$.
 Fig.(\ref{fig:diffopCom_2.5}) plots K-complexity at all time scales for $\Lambda=2.5$ with $N=5$, and $M=3$. The linear growth and the saturation can be observed at times of $\mathcal{O}(D)$, and at $\mathcal{O}(D^{2})$ respectively. For $N=5$, and $M=3$, $D=441$ and $K\sim 421$.}
\label{fig:FComFLan2535}
\end{figure}

{\color{black}In this section, we provide the numerical results obtained from studying the Lanczos coefficients and Krylov complexity for a choice of the initial operator that is different from the initial operator considered in the main text. We work in $M=3$ and $N=5$ system. The appropriate hermitian operator that we choose for numerical computations is given by the number operator at the first and second sites: }
\begin{eqnarray}
\label{iniopdiff}\mathcal{O}_{\text{in}}=\hat{n}_{1}+\hat{n}_{2}
\end{eqnarray}
{\color{black}The numerical results obtained are shown in the Fig (\ref{fig:FComFLan2535}). One can take other choices of different hermitian operators that do not commute with the Bose-Hubbard Hamiltonian \eqref{hamlt}. For instance, the choice of the initial operator where the number operator at all sites is taken commute with the Hamiltonian \eqref{hamlt} and thus makes the Lanczos algorithm trivial.}
\section{Numerical fits of K-complexity at different time scales}
\label{A2}
{\color{black}Below we discuss the numerical fits done to obtain the different behaviours of $C_{k}(t)$ at different time scales. We have shown a case where $\Lambda=0.1$, $N=5$, and $M=3$. We found that the exponent $\alpha=0.49635$ controlling the exponential behaviour of K-complexity at early times is approximately equal to the slope of $b_{n}$, $\beta=0.53538$ in the linear regime as was expected.}
\vspace{1cm}
\begin{figure}[h!]
\begin{subfigure}{0.47\linewidth}
\centering
 \includegraphics[width=\linewidth]{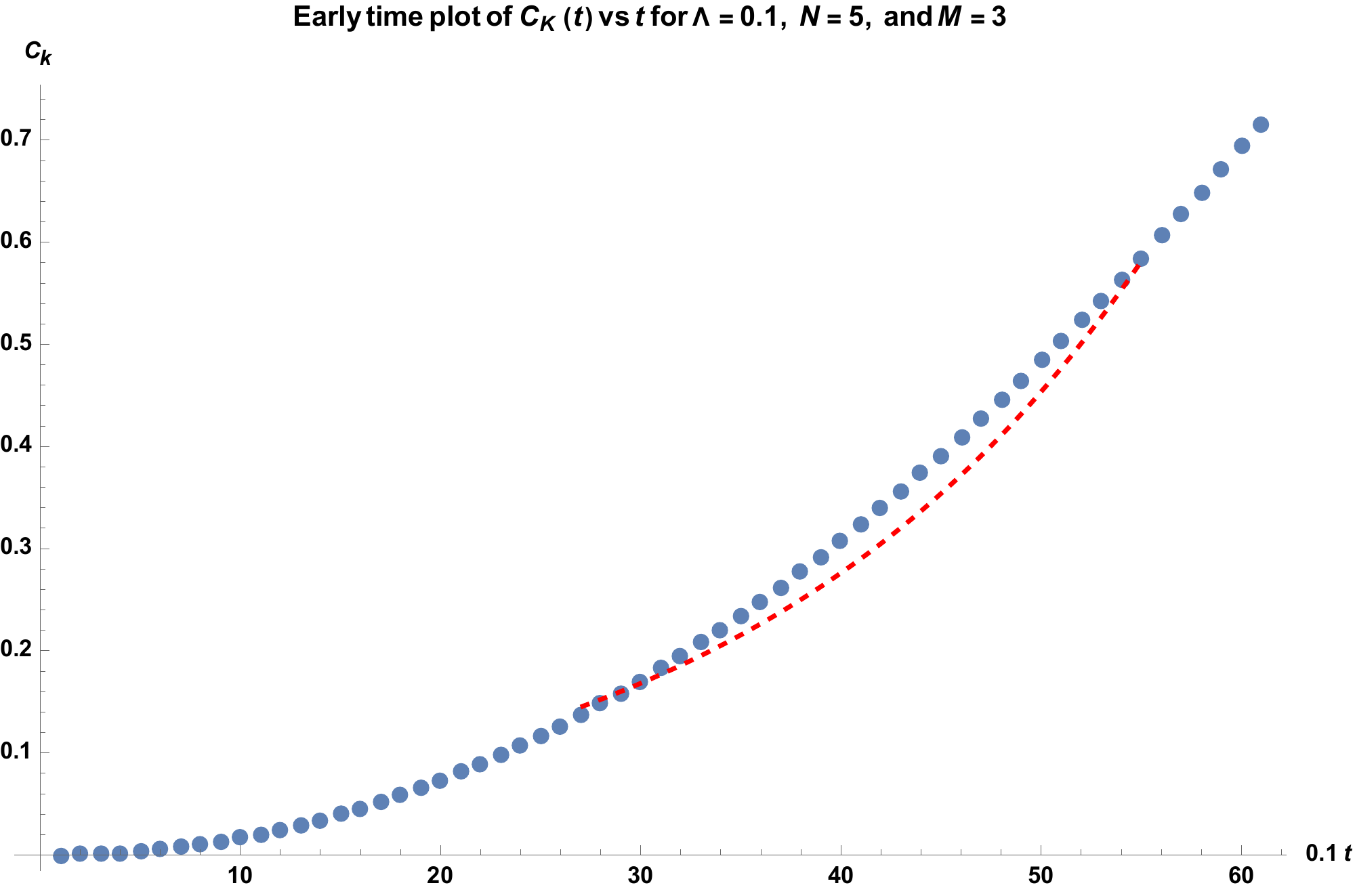}
 \caption{}
 \label{fig:numexpfit}
 \end{subfigure}
 \begin{subfigure}{0.47\linewidth}
  \includegraphics[width=\linewidth]{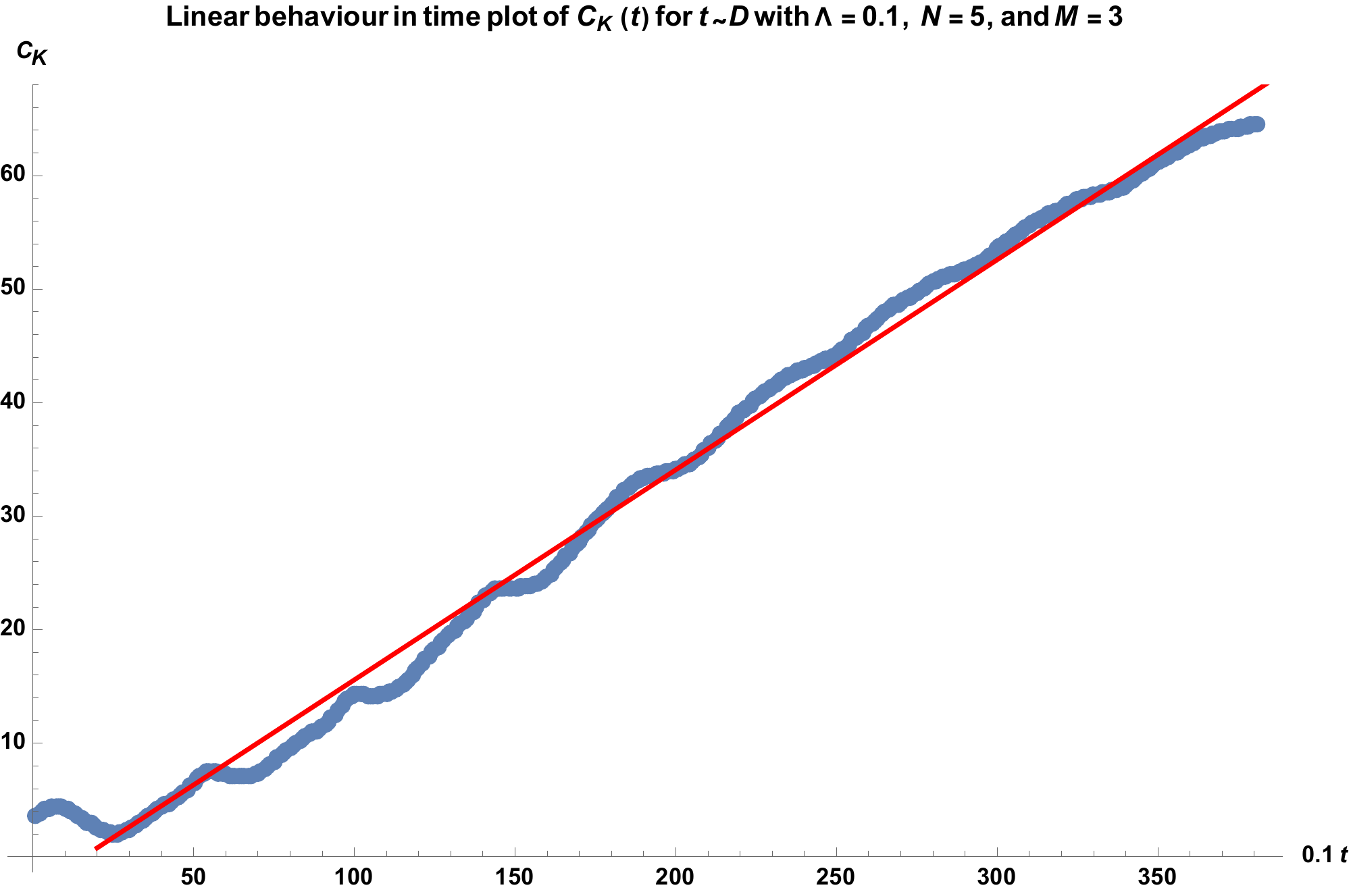}
  \caption{}
  \label{fig:numlinfit}
   \end{subfigure}
 \caption{\sf Plot of K-complexity at all different scales for $\Lambda=0.1$ with $N=5$, and $M=3$. Numerical fits are included at both the early time plot in Fig.(\ref{fig:numexpfit}) near the scrambling time $t_{s}=0.48$ and at times $\sim \mathcal{O}(D)$ in 
 Fig.(\ref{fig:numlinfit}) respectively. Fig.(\ref{fig:numexpfit}) shows the exponential growth of K-complexity at early times ($t\sim \log(ln D)$). The non-linear fit function is $C_{K}(t)=0.0379053\,\exp{(0.49635\,t)}$ with $r^{2}=0.989944$ denoted by  \textbf{red dashed lines}. Fig.(\ref{fig:numlinfit}) shows the linear growth of K-complexity at times of $\mathcal{O}(D)$.The linear fit function is given by $C_{K}(t)=-2.92944 + 1.85039\,t$ with $r^{2}=0.992016$ denoted by \textbf{red line}.}
\label{fig:numfit}
\end{figure}

\section{More on the operator growth and Lanczos coefficients} \label{B}
We give an example of the growth of the operator in the Krylov basis. Given the Hamiltonian in \ref{BHHam}, and the general form of the initial operator in \ref{iniop}, we examine the steps of the Lanczos algorithm. We use the notion of Frobenius inner product defined in \ref{defnorm} and the bosonic commutation relation is given by
\begin{eqnarray}
    [a_{n},a^{\dag}_{n}]=1\,,
\end{eqnarray}
where the subscript $n$ represents the site number. We start with the three-particle and three-site Bose-Hubbard model where $\Lambda= 2.5$, and the initial operator is given by
\begin{eqnarray}
\label{iniop1}   \mathcal{O}_{ini}=\hat{n}_{1}+\hat{n}_{2}-i(a^{\dag}_{1}a_{2}-a^{\dag}_{2}a_{1}) 
\end{eqnarray}
where $\hat{n}_{1}=a^{\dag}_{1}a_{1}$ and $\hat{n}_{2}=a^{\dag}_{2}a_{2}$ are the number operators on the 1st and 2nd site respectively.
\begin{itemize}

\item $\mathcal{O}_{0}$: The operator $\mathcal{O}_{ini}$ is not normalised. We normalise it using the inner product defined in \ref{defnorm} to get our starting operator $\mathcal{O}_{0}$ as:
\begin{eqnarray}
|\mathcal{O}_{0})=\frac{1}{||\mathcal{O}_{ini}||}|\mathcal{O}_{ini})= 0.353  |\mathcal{O}_{ini}).
\end{eqnarray}

\item $\mathcal{O}_{1}$: Computing $\mathcal{O}_{1}$ needs the evaluation of $\mthA_1$:
\begin{eqnarray}
\mthA_1=[H,\mathcal{O}_{0}]\,.
\end{eqnarray}

Next, we perform reorthogonalization on $\mthA_1$ by writing
\begin{eqnarray}
|\mthA_1)\longrightarrow |\mthA_1)-|\mthO_0)(\mthO_0|\mthA_1)
\end{eqnarray}
and get
\begin{eqnarray}
\mthA_1 &=& 0.707i ( \hat{n}_{2}-\hat{n}_{1}) + 0.353(a^{\dag}_{2}a_{3}-a^{\dag}_{3}a_{2}-i\,a^{\dag}_{1}a_{3}-i\,a^{\dag}_{3}a_{1})+\nonumber \\
&& 0.884\, i\,(a^{\dag}_{2}a^{\dag}_{1}a_{2}a_{2} + a^{\dag}_{2}a^{\dag}_{2}a_{2}a_{1}-a^{\dag}_{1}a^{\dag}_{1}a_{2}a_{1}-a^{\dag}_{2}a^{\dag}_{1}a_{1}a_{1})\,.
\end{eqnarray}
The first Lanczos coefficient is:
\begin{eqnarray}
b_{1}= \sqrt{(\mthA_1|\mthA_1)}=2.574.
\end{eqnarray}
The matrix representation of the operators $\mthA$ is written in the $D=10$ dimensional Hilbert space. As a result, at this level of iteration, the orthogonalized Krylov operator is:
\begin{eqnarray}
\mathcal{O}_{1}=\frac{1}{b_{1}}\,|\mthA_1)&=& 0.275\,i ( \hat{n}_{2}-\hat{n}_{1}) + 0.137(a^{\dag}_{2}a_{3}-a^{\dag}_{3}a_{2}-i\,a^{\dag}_{1}a_{3}-ia^{\dag}_{3}a_{1})+\nonumber \\
&& 0.343\, i\,(a^{\dag}_{2}a^{\dag}_{1}a_{2}a_{2} + a^{\dag}_{2}a^{\dag}_{2}a_{2}a_{1}-a^{\dag}_{1}a^{\dag}_{1}a_{2}a_{1}-a^{\dag}_{2}a^{\dag}_{1}a_{1}a_{1})\,.
\end{eqnarray}
Note that, the operator $\mathcal{O}_{0}$ is a double-site operator. It grows to $\mathcal{O}_{1}$, which is a mix of double-site and quadruple-site operators. At the next step, the operator $\mathcal{O}_{1}$ will grow to a six-site operator contribution to give $\mathcal{O}_{2}$. However, the operators cannot grow more than a six-site contribution because of the restriction on the total number of particles $N=3$. Thus, the operator remains as a six-site operator until it exhausts all the linearly independent contributions.
The rest of the orthonormal operators of the Krylov basis can be generated using the same approach. In this case, the algorithm terminates at the $n=92$ level of iteration giving $b_{92}=0$. Hence the Krylov dimension is $K=91$ which exactly saturates the bound mentioned in \ref{Kbasisbound}. For a larger value of $N$, the operator growth does not stop at a six-site operator. Hence, the analysis becomes more and more tedious and can be performed by a numerical analysis owing to the results in the main text.
\end{itemize}
\bibliographystyle{JHEP}
\bibliography{main}

\end{document}